\documentclass[a4paper,preprintnumbers,floatfix,twocolumn,aps,prb,unsortedaddress,superscriptaddress]{revtex4-2}

\usepackage[para]{threeparttable} 
\usepackage{amsmath}
\usepackage{graphicx} 
\usepackage{booktabs}
\usepackage{subcaption}
\usepackage{url}
\usepackage{hyperref}
\usepackage{miller}
\usepackage{bm}
\usepackage{enumitem}
\usepackage{color}

\usepackage[paperwidth=210mm,paperheight=297mm,centering,hmargin=1.7cm,vmargin=2.5cm]{geometry}

\usepackage[utf8]{inputenc}

\begin{document}

\title{Simple machine-learned interatomic potentials for complex alloys}

\author{J. Byggmästar}
\thanks{Corresponding author}
\email{jesper.byggmastar@helsinki.fi}
\affiliation{Department of Physics, P.O. Box 43, FI-00014 University of Helsinki, Finland}
\affiliation{Finnish Center for Artificial Intelligence, FCAI}
\author{K. Nordlund}
\affiliation{Department of Physics, P.O. Box 43, FI-00014 University of Helsinki, Finland}
\author{F. Djurabekova}
\affiliation{Department of Physics, P.O. Box 43, FI-00014 University of Helsinki, Finland}
\affiliation{Helsinki Institute of Physics, Helsinki, Finland}

\date{\today}

\begin{abstract}
{Developing data-driven machine-learning interatomic potentials for materials containing many elements becomes increasingly challenging due to the vast configuration space that must be sampled by the training data. We study the learning rates and achievable accuracy of machine-learning interatomic potentials for many-element alloys with different combinations of descriptors for the local atomic environments. We show that for a five-element alloy system, potentials using simple low-dimensional descriptors can reach meV/atom-accuracy with modestly sized training datasets, significantly outperforming the high-dimensional SOAP descriptor in data efficiency, accuracy, and speed. In particular, we develop a computationally fast machine-learned and tabulated Gaussian approximation potential (tabGAP) for Mo--Nb--Ta--V--W alloys with a combination of two-body, three-body, and a new simple scalar many-body density descriptor based on the embedded atom method.}
\end{abstract}

\maketitle

\section{Introduction}
\label{sec:intro}

Machine-learning (ML) interatomic potentials are now popular tools to achieve near-quantum accuracy in classical molecular dynamics simulations~\cite{behler_perspective_2016,deringer_gaussian_2021,Mis21}. The development of new ML frameworks with different ways of quantifying the local atomic environments with atomistic descriptors has been rapid and still continues. The typical starting point for developing a new type of ML potential is to demonstrate good accuracy for single-element materials with structural diversity (e.g. different phases or various crystallographic defects), molecules, or fixed-composition alloys and compounds~\cite{behler_generalized_2007,bartok_gaussian_2010,thompson_spectral_2015,drautz_atomic_2019,fan_neuroevolution_2021}. This has left the challenges of developing ML potentials for materials of multiple elements much less explored, especially potentials that are transferable to arbitrary compositions of the included elements~\cite{byggmastar_modeling_2021,li_complex_2020,hajinazar_stratified_2017-1,rosenbrock_machine-learned_2021,nikoulis_machine-learning_2021}. The main challenge for materials with many elements is the vast composition space that needs to be sampled by the training data. For a general and transferable ML potential, the combined configuration space of both chemical and structural diversity of the material must be sampled.

The structures that make up the training data are typically computed within density functional theory (DFT) to obtain accurate total energies and forces. The size of the training dataset of structures must be kept at a tractable level, partly due to the computational cost of DFT, but also due to practical computer memory limits. For multicomponent materials, this calls for a data-efficient ML potential. The success heavily relies on how well the ML potential can interpolate between chemical compositions. A counterproductive fact to this is that the dimensionality of many commonly used many-body descriptors scales poorly with the number of elements, so that the descriptor space that the training data should cover becomes enormous~\cite{artrith_efficient_2017,imbalzano_automatic_2018,willatt_feature_2018,darby_compressing_2021}. This issue was discussed and addressed for the SOAP descriptor~\cite{bartok_representing_2013} in a recent article~\cite{darby_compressing_2021}, where the dimensionality scaling is reduced from quadratic to linear with respect to the number of elements.

These challenges are particularly demanding when developing ML potentials for high-entropy alloys, which is an intensely studied class of materials that typically contain five or more elements with significant concentrations in random solid solutions~\cite{yeh_nanostructured_2004,cantor_microstructural_2004,senkov_refractory_2010,tsai_high-entropy_2014}. To our knowledge, our previous work~\cite{byggmastar_modeling_2021} is to date the only ML potential for a five-element alloy system (Mo--Nb--Ta--V--W) that is general enough to be applicable to the entire range of compositions, and from single crystals to defects and the liquid phase. Due to the novelty of the work, the training database was constructed in a somewhat tedious trial-and-error manner until good testing errors and material properties were achieved. In the process of generating the training database, we found that a typical Gaussian approximation potential (GAP)~\cite{bartok_gaussian_2010} with the high-dimensional SOAP descriptor never reached satisfactory accuracy when incrementally adding more training data. Instead, we found that using simple low-dimensional two-body and three-body descriptors in a GAP was enough to outperform a SOAP-GAP for crystalline alloys. This came with the significant added bonus that a pure 2+3-body potential can be tabulated and evaluated efficiently using cubic-spline interpolations (which we called tabGAP), leading to a speed-up of two orders of magnitude~\cite{byggmastar_modeling_2021,glielmo_efficient_2018,vandermause_--fly_2020}.

Motivated by these findings and the above-mentioned challenges, the aim of this work is to systematically study and compare the learning curves and achievable accuracies of ML potentials for many-element materials with different combinations of descriptors. Specifically, we wish to answer the following questions when developing many-element ML potentials, using the five-element Mo--Nb--Ta--V--W system as a case study: (1) how do various computationally fast low-dimensional descriptor combinations compare with more expensive high-dimensional descriptors? (2) how much training data does one need for a given set of descriptors? and (3), with a given practically feasible amount of training data and computational speed in consideration, what is the optimal set of descriptors? 


\section{Methods}
\label{sec:methods}

\subsection{Machine-learning potentials}
\label{sec:mlpot}

We use the Gaussian approximation potential (GAP) framework as implemented in the \textsc{quip} code~\cite{bartok_gaussian_2010,QUIP}, with an external fixed analytical screened Coulomb potential for the short-range repulsion~\cite{byggmastar_machine-learning_2019}. GAP uses Gaussian process regression to establish a prediction for the energy, which can be written as a sum over kernels $K$ weighted linearly by the optimised regression coefficients $\alpha_s$. An appealing feature of GAP is that multiple different descriptors can be used in separate terms and simultaneously trained as one model. This makes direct comparison of different combinations of descriptors easy without changing the underlying ML regression. With the external repulsive potential $E_\mathrm{rep.}$, the total energy of a system of atoms can in general then be expressed as
\begin{equation}
    E_\mathrm{tot.} = E_\mathrm{rep.} + \sum_d \delta^2_d \sum_{i} \sum_s \alpha_{s} K_d (\bm{q}_{d, i}, \bm{q}_{d, s}).
    \label{eq:GAP}
\end{equation}
Here, $d$ denotes a given descriptor, $i$ runs over all descriptor environments in the system, and the final sum over $s$ is the sparse Gaussian process regression sum that runs over a selected set of descriptor environments from the training data. $E_\mathrm{rep.}$ is the screened Coulomb potential, $\delta^2_d$ is a weighting prefactor for the descriptor $d$, $K_d$ is the kernel function used for descriptor $d$, and $\bm{q}_{d, i}$ the descriptor output for the local atomic environment $i$. For more details about GAP and the repulsive potential, we refer to previous work~\cite{bartok_gaussian_2015,bartok_machine_2018,byggmastar_machine-learning_2019}.

\subsection{Descriptors}
\label{sec:desc}

In all trained potentials, we always use a pair descriptor (``2b''), i.e. the interatomic distance, to get a machine-learned pair-potential contribution. Together with the screened Coulomb pair potential $E_\mathrm{rep.}$ they provide a good basis that accounts for accurate pair repulsion and a rough cohesive interaction. Beyond 2b, we combine additional descriptors of different complexity. These include the three-body (``3b'') descriptor $[(r_{ij}+r_{ik}), (r_{ij}-r_{ik})^2, r_{jk}]$~\cite{bartok_gaussian_2015} (where e.g. $r_{ij}$ is the distance between atoms $i$ and $j$), the smooth overlap of atomic positions (SOAP) descriptor~\cite{bartok_representing_2013}, one of the recently introduced compressed SOAP versions (cSOAP)~\cite{darby_compressing_2021}, and a simple scalar density descriptor (``EAM'') based on the embedded atom method~\cite{daw_embedded-atom_1984,finnis_simple_1984} which we introduce in detail below. The compression of the full SOAP descriptor in cSOAP is beneficial for many-element materials, as the scaling of descriptor vector length with number of elements is reduced from quadratic to linear. Here we include only the $\nu, \mu=1, 1$ compression discussed in~\cite{darby_compressing_2021}, as it provides the best accuracy.

We introduced and used the EAM descriptor previously for pure Fe~\cite{byggmastar_multiscale_2022} (inspired by Ref.~\cite{zeni_gaussian_2020}) and showed that it, despite its simplicity, provides a significant boost in accuracy at negligible computational cost when combined with the 2b and 3b descriptors. In the following, we generalise it for many-element interactions. The EAM descriptor for the local environment of an atom $i$ surrounded by neighboring atoms $j$ is the total pairwise-contributed scalar density, just like in conventional EAM potentials:
\begin{equation}
    \rho_i = \sum_{j \neq i} \kappa_{ij} \varphi(r_{ij}).
\end{equation}
Here $\kappa_{ij}$ is a prefactor that depends on the chemical species of $i$ and $j$ and $\varphi (r)$ is a scalar function for the pair-density contribution. Compared to conventional EAM potentials, the machine-learning task when using the EAM descriptor is simply the fitting of an embedding function for each element (while simultaneously including the contributions of possible other descriptors to the total energy).

The pairwise density contributions given by $\varphi(r)$ can in principle be any function that smoothly reaches zero at the cutoff distance. It could also be a freely adjustable spline-like function like in many conventional EAM potentials~\cite{mendelev_development_2003}. However, since it is part of the descriptor and must be treated as a set of hyperparameters that need to be fixed before the ML training, in this case it would have to be pre-fitted or adjusted with a hyperparameter-optimisation strategy. Here, we instead use simple analytical functions for $\varphi(r)$, favouring functions that reach finite values as $r \longrightarrow 0$ so that the strongly repulsive range is well-behaved. We have currently implemented three different functions (listed and illustrated in Appendix~\ref{sec:appendix}). One is a simple smoothly decaying polynomial function of arbitrary degree, and the other two are functions that are often used as cutoff functions for potentials and descriptors. The latter two can be shifted to reach a constant value for distances $r \leq r_\mathrm{min}$, where $r_\mathrm{min}$ is user-defined. Setting $r_\mathrm{min}>0$ can be useful so that the ML energy contribution from the EAM descriptor is forced to a constant (and the force to zero), because at short distances the external screened Coulomb potential (calculated with separate DFT calculations) should dominate the interaction~\cite{ziegler_stopping_1985,nordlund_repulsive_1997}.

Just like in normal EAM-like potentials, the pair-density contribution of atom $j$ of element $B$ to the position of atom $i$ (element $A$) can depend: (1) only on the element of $j$; $\kappa_{AB} = \kappa_B$, like in the original EAM alloy potentials~\cite{foiles_embedded-atom-method_1986}, (2) on the element pair $AB$ symmetrically; $\kappa_{AB} = \kappa_{BA}$, as is typical for Finnis-Sinclair-like potentials~\cite{finnis_simple_1984,ackland_development_2004}, or (3) asymmetrically $\kappa_{AB} \neq \kappa_{BA}$. In addition to the trivial case of $\kappa_{AB} = 1$, where the descriptor is blind with respect to the elements of the neighbouring atoms, we have implemented these options as functions of the atomic numbers $Z_A$, $Z_B$ as:
\begin{equation}
   \kappa_{AB} =
   \begin{cases}
    1 \\
    \sqrt{Z_B}/10 \\
    \sqrt{Z_A Z_B} / 40 \\
    Z_A^{0.1} \sqrt{Z_B}/10 \\
   \end{cases}
   \label{eq:prefactor}
\end{equation}
Square roots are taken to limit the dependence on very different atomic numbers and the division by 10 or 40 is purely out of convenience to produce similar values of $\kappa$. The corresponding names in the code are \texttt{blind}, \texttt{EAM}, \texttt{FSsym}, and \texttt{FSgen}. One could also implement ways to freely optimise the values of $\kappa_{AB}$ for the involved elements instead of the above fixed dependencies on atomic numbers (although it is unclear how much it would help with the model accuracy given the simplicity of the descriptor).

$\varphi (r)$ and $\kappa_{AB}$ can be chosen independently of each other from the above options. Note that multiple EAM descriptor terms with different $\varphi (r)$ and $\kappa_{AB}$ can be used for the same element in GAP, which can be seen as a generalisation of the two-band EAM potential~\cite{ackland_two-band_2003} or as analogous to using multiple radial functions in high-dimensional many-body descriptors~\cite{behler_generalized_2007}.

For comparison of accuracy, data efficiency, and computational speed, we choose to test the following combinations of descriptors: 2b only, 2b+EAM, 2b+$N$EAM, 2b+3b, 2b+3b+EAM, 2b+SOAP, and 2b+cSOAP. $N$EAM refers to using multiple ($N$) EAM descriptors per element. Note that we neglect combinations such as 2b+3b+$N$EAM, 2b+EAM+SOAP, and 2b+3b+SOAP because they provide similar accuracy to some of the above combinations but at much higher computational cost. Additionally, for reasons discussed in the next section, we focused on comparing all reasonable combinations of low-dimensional descriptors (such as 2b+3b+EAM) to a few high-dimensional GAPs such as 2b+SOAP.

\subsection{tabGAP: tabulated Gaussian approximation potentials}
\label{sec:tabgap}

As discussed in more detail in our previous work~\cite{byggmastar_modeling_2021,byggmastar_multiscale_2022}, a major advantage of using only low-dimensional descriptors (2b, 3b, EAM) is that the corresponding GAP energy predictions can be mapped onto discrete grids for each descriptor~\cite{glielmo_efficient_2018}. Interpolating between these grid points with cubic splines is significantly faster than carrying out the Gaussian process regression in Eq.~\ref{eq:GAP}. The 2b energies simply become 1D spline-interpolated pair potentials for each element pair. The 3b descriptor is a vector in three dimensions~\cite{bartok_gaussian_2015}, so the 3b energies must be mapped onto 3D grids for each element triplet. The EAM energies become a conventional tabulated EAM potential file with 1D splines for the pair density function and embedding energies. We refer to the tabulated version of a low-dimensional GAP as a tabGAP. For example, a 2b+3b+EAM GAP becomes the tabGAP:
\begin{equation}
\begin{aligned}
    & E_\mathrm{GAP\text{-}2b+3b+EAM} \overset{\mathrm{tab.}}{\simeq} E_\mathrm{tabGAP} =\\
    & \sum_{i < j}^N S_{\mathrm{rep. + 2b}}^\mathrm{1D} (r_{ij}) + \sum_{i, j<k}^N S_{ijk}^\mathrm{3D} (r_{ij}, r_{ik}, \cos \theta_{ijk}) \\
    & + \sum_{i}^N S_\mathrm{emb.}^\mathrm{1D} \left(\sum_j^N S_\varphi^\mathrm{1D} (r_{ij}) \right),
\end{aligned}
\end{equation}
where $S$ represent the 1D and 3D cubic-spline interpolations. The 3D spline interpolates between points on a $(r_{ij}, r_{ik}, \cos \theta_{ijk})$ grid~\cite{byggmastar_modeling_2021}, where $\theta_{ijk}$ is the angle between the $ij$ and $ik$ interatomic bonds.

Our code for generating tabGAP potential files and using them in \textsc{lammps}~\cite{thompson_lammps_2022} is available from~\cite{tabgap}.

\subsection{Learning curves and training and testing data}
\label{sec:data}

For investigating the learning curves of the different combinations of descriptors for many-element materials, we use the Mo--Nb--Ta--V--W training and testing data from~\cite{byggmastar_modeling_2021}. The dataset consists of various crystalline and liquid structures spanning the entire range of compositions from the pure metals to five-element alloys. The actual training data are the total energies, forces, and for some structures virial stresses computed with density functional theory using the \textsc{vasp} code~\cite{kresse_efficient_1996} (see details in~\cite{byggmastar_modeling_2021}). The training dataset includes subsets of the pure-element training data from~\cite{byggmastar_gaussian_2020}, finite-temperature randomly ordered and ordered single-crystals of the entire alloy composition range, defects (vacancies and self-interstitial atoms) in the equiatomic MoNbTaVW alloy, liquid structures for various alloy compositions, surfaces, dimers, and structures with short interatomic separations for fitting repulsion. The training structures are described in more detail in the Supplemental document of Ref.~\cite{byggmastar_modeling_2021} and is available from Ref.~\cite{byggmastar_ida_data_2022}.

To study the learning rates, we generate different sized subsets of the full training dataset by picking increasingly many structures (from 2\,\% to 100\,\%). The structures are picked (almost) randomly as follows. First, since we are mainly interested in the learning curves of alloy structures, all pure-element training structures are kept as a basis for all subsets. Second, in the remaining alloy structures, the random sampling is done separately for each class of structures (\texttt{config\_type} in GAP language), so that e.g. 10\% each of the single-crystal random alloys, ordered alloys, liquids, vacancy structures, etc. are picked separately. Three random subsets are generated for each training data size to assess the statistical differences of the random picking. All reported test and train errors are the averages of the resulting three potentials per training data subset.

The full training dataset contains 2859 structures with total energies and in total 146101 atoms, each with three force components. 1614 structures (105535 atoms) of these are the alloy structures from which different fractions are picked as described above. The remaining 1245 structures make up the pure-element data that are always included.

The testing data is the same for all training runs and contains single-crystal alloys of random compositions and equiatomic MoNbTaVW, ordered alloys, interstitials and vacancies in equiatomic MoNbTaVW, and liquid MoNbTaVW. When studying the learning curves, we compute the root-mean-squared errors (RMSE) of the training and testing data. For both training and testing data, the structures are grouped into ``crystals'' and ``liquids'', where ``crystals'' contain all crystalline (body-centred-cubic) structures, including structures with defects. The RMSEs are calculated separately for these two groups. Together, they provide a good measure how well a potential is able to model both high-symmetry lattice structures and completely disordered structures.

The crystal test set contains in total 325 structures (13520 atoms in total) and the liquid test set 11 structures (1408 atoms). Because the liquid test set only includes 11 total energies, we sometimes only report the force errors for liquids as they provide a more statistically reliable test error (4224 force components). Both the training and testing data are available from Ref.~\cite{byggmastar_ida_data_2022}.

\subsection{Hyperparameters}

\begin{figure}
    \centering
    \includegraphics[width=\linewidth]{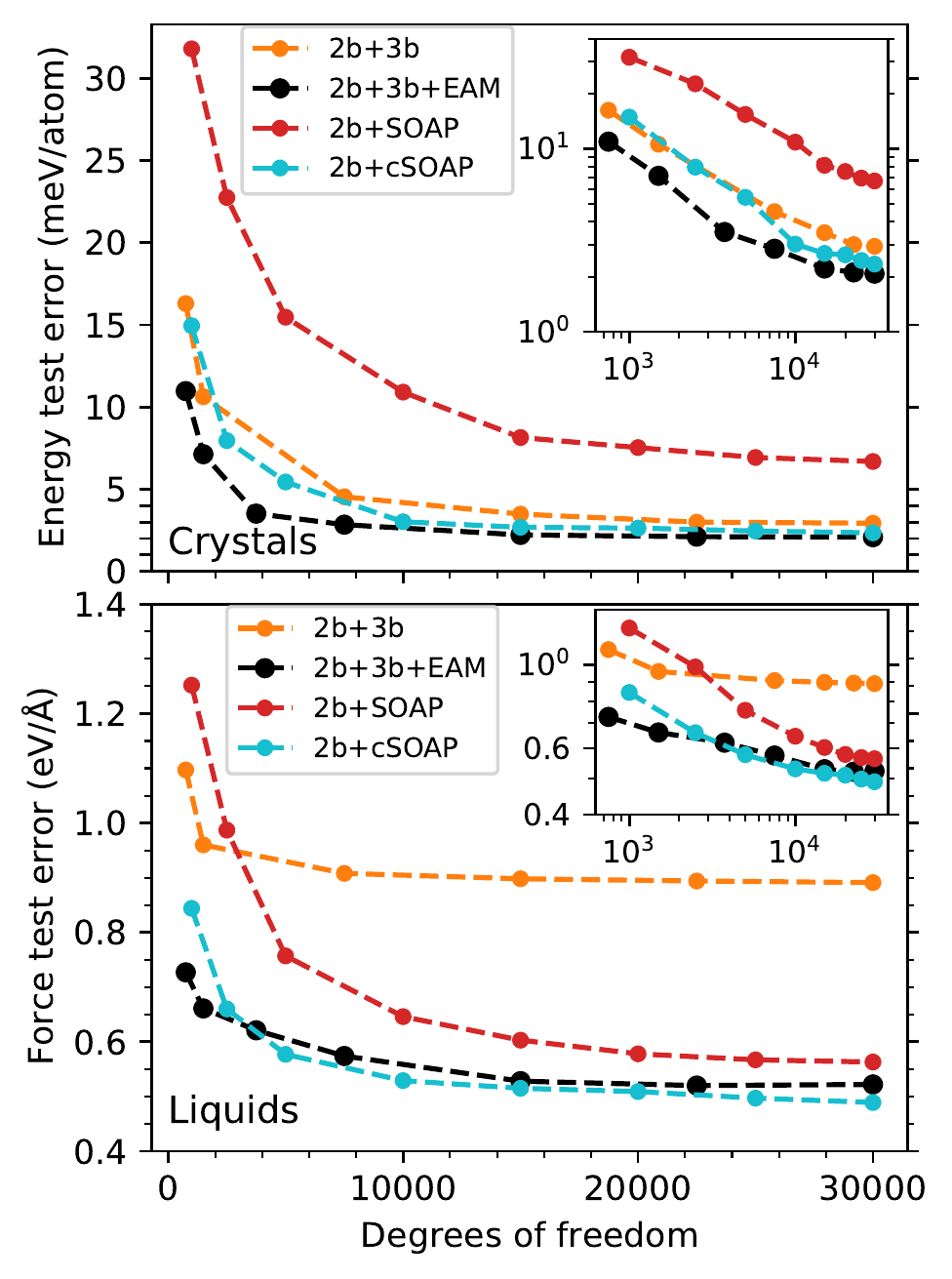}
    \caption{Learning curves as functions of the total number of degrees of freedom for potentials using the 3b and SOAP descriptors, showed separately for the crystal alloy test set (top) and the liquid alloy test set (bottom). The insets show the same data on logarithmic scales.}
    \label{fig:nsparse}
\end{figure}

The necessary hyperparameters of GAP and the different descriptors are similar as in our previous work~\cite{byggmastar_modeling_2021}. In short, the cutoff distances are 5 Å for the simplest and computationally fastest 2b and EAM descriptors and 4.1 Å for 3b, SOAP, and cSOAP (in Ref.~\cite{byggmastar_modeling_2021} we used 5 Å also for 3b). $n_\mathrm{max}$ and $l_\mathrm{max}$ for the radial and spherical harmonics expansion of SOAP were 8 and 6, respectively. The choice is based on a balance between sufficient accuracy and manageable memory requirement (which for five elements is substantial, up to the maximum 1.5 TB available per node in our computer cluster).
The regularisation parameters for energies and forces in GAP~\cite{bartok_gaussian_2015} are the same for all combinations of descriptors. In short, high accuracy for crystals is emphasized by using five times stronger weighting factors than for liquids. All input hyperparameters and necessary input files are provided in Ref.~\cite{byggmastar_ida_data_2022} for reproducibility.

Perhaps the most critical hyperparameter is the number of sparse descriptor environments picked from the training data for the Gaussian process regression ($N_\mathrm{sparse}$ in GAP language~\cite{bartok_gaussian_2015}). $N_\mathrm{sparse}$ effectively defines the number of degrees of freedom of the fit and should be converged to a suitable value for each descriptor. Since we have different-sized training sets, one should in principle increase the degrees of freedom as a function of the training data size, to optimise the computational speed of the potential and to avoid under- or overfitting. However, we found that using the same values for all training data sizes does not affect the training and testing errors significantly and is therefore appropriate for our main purpose of studying learning curves. The convergence of the test errors as a function of number of degrees of freedom is shown in Fig.~\ref{fig:nsparse} for the 3b- and SOAP-containing combinations of descriptors trained to the full training dataset. The scalar-valued 2b and EAM descriptors always have the same 20 degrees of freedom, so that Fig.~\ref{fig:nsparse} only shows the convergence of the 3b and SOAP descriptors. From Fig.~\ref{fig:nsparse}, we choose $N_\mathrm{sparse}=300$ for the 3b descriptor and $N_\mathrm{sparse}=4000$ for SOAP, corresponding to comparable total numbers of degrees of freedom ($75 \times 300 = 22500$ for 3b, since with five elements there are 75 unique element triplets, and $5 \times 4000 = 20000$ for SOAP).

\begin{figure}
    \centering
    \includegraphics[width=\linewidth]{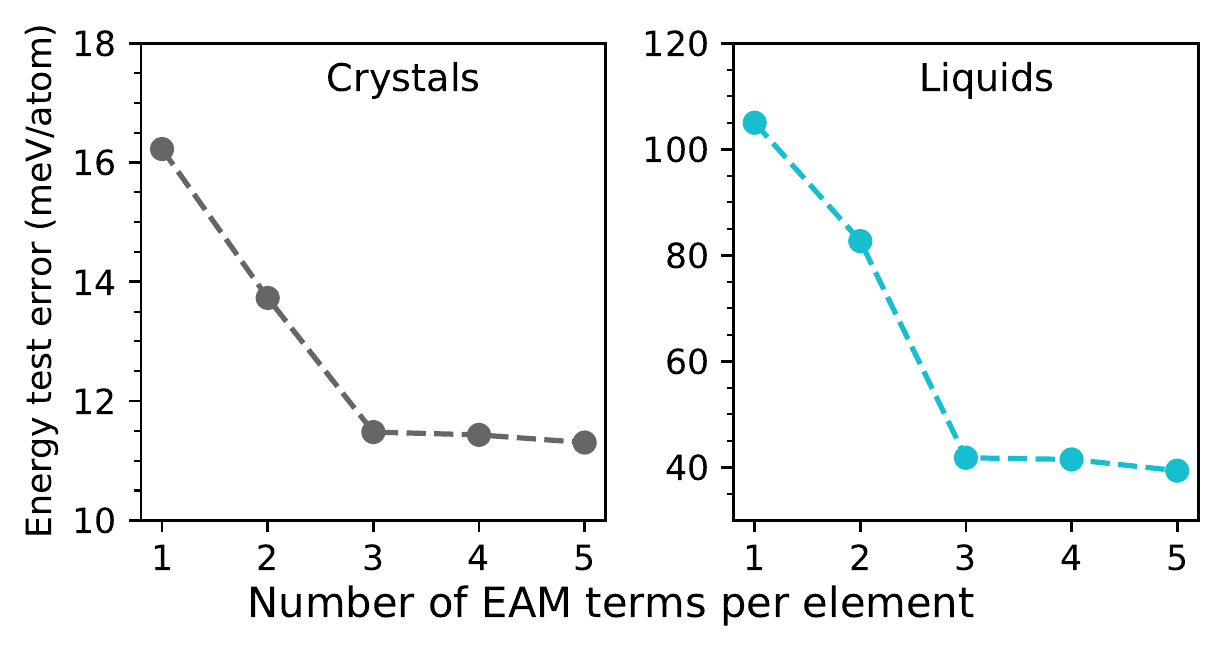}
    \caption{Convergence of the test errors as a function of the number of embedding terms for each element (i.e. number of EAM density descriptors) in the 2b+$N$EAM potential. Three EAM terms were used in all other 2b+$N$EAM results.}
    \label{fig:nband}
\end{figure}

For the EAM descriptor discussed above, there are several hyperparameters. We found that all options of Eq.~\ref{eq:prefactor} for the prefactor $\kappa_{AB}$ produce similar results. Somewhat surprisingly, the ``blind'' option $\kappa_{AB}=1$ proved to be slightly better than the others. This is likely because most Mo--Nb--Ta--V--W alloys have mixing enthalpies very close to zero~\cite{byggmastar_modeling_2021}, so that differentiating between elements in the pair-density contributions is not critical (but note that the machine-learning ``embedding energy'' is still unique to each element). Hence, we set $\kappa_{AB}=1$ for all EAM descriptors. For the 2b+$N$EAM potential, we also need to choose the number of EAM descriptors per element, $N$. Fig.~\ref{fig:nband} shows the convergence of the test errors as functions of $N$, using multiple pair density functions (Eq.~\ref{eq:pairdens3}) with different values of $r_\mathrm{min}$. Combining different pair density functions (Eq.~\ref{eq:pairdens1}--\ref{eq:pairdens3}) and $\kappa_{AB}$ leads to very similar results. Fig.~\ref{fig:nband} shows a rapid convergence at $N=3$, which we therefore use for the 2b+$N$EAM potentials. The difference in accuracy compared to using only one EAM term like in normal EAM potentials is significant, especially for liquids (Fig.~\ref{fig:nband}).

\section{Results and discussion}
\label{sec:res}

\subsection{Learning curves}

\begin{figure*}
    \centering
    \includegraphics[width=\linewidth]{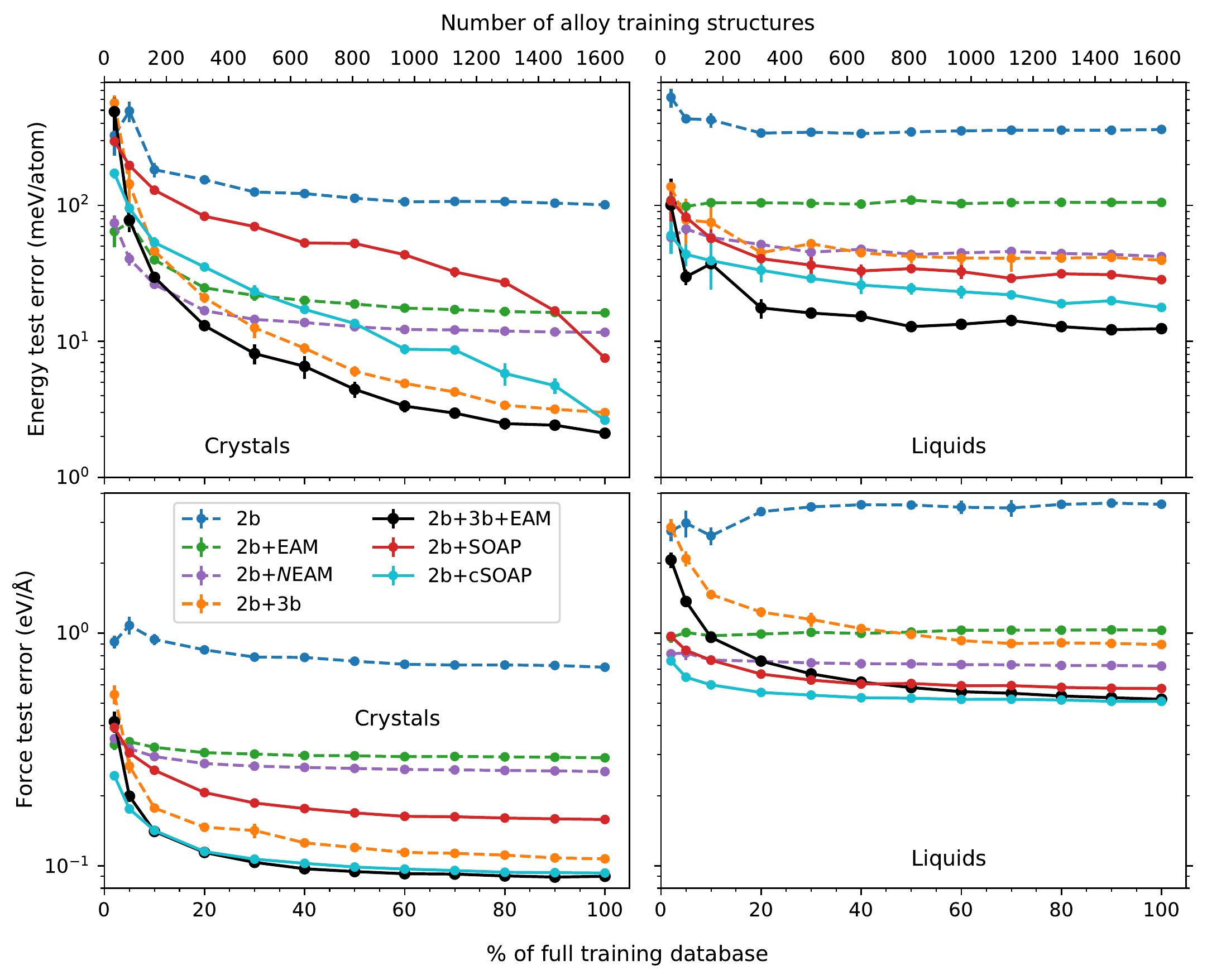}
    \caption{Learning curves for the different combinations of descriptors. Test errors are shown as functions of the size of the training set of Mo--Nb--Ta--V--W alloys, given as fractions of the full alloy training dataset or number of structures (top axis). A training dataset of size zero contains only pure-element structures. The learning curves are showed separately for energy and force test errors (RMSE) and for the crystal and liquid test sets.}
    \label{fig:learning}
\end{figure*}

Figs.~\ref{fig:nsparse} and~\ref{fig:learning} show the main results of the learning curves using the different combinations of descriptors. As briefly discussed above, Fig.~\ref{fig:nsparse} shows the test errors as functions of the total number of degrees of freedom for the 3b and SOAP-like descriptor terms of the potentials. Both Figs.~\ref{fig:nsparse} and \ref{fig:learning} show that the low-dimensional 2b+3b+EAM combination of descriptors reaches the best accuracy. Remarkably, it outperforms 2b+SOAP for both crystals and liquids. 2b+3b also outperforms 2b+SOAP, as we previously discovered in Ref.~\cite{byggmastar_modeling_2021}, but only for crystalline structures. The improvement when including the EAM descriptor in 2b+3b+EAM compared to 2b+3b is significant, especially for liquids. The compressed SOAP version also provides a significant improvement over the full SOAP descriptor and the overall accuracy of 2b+cSOAP is almost identical to 2b+3b+EAM when using the full training dataset.

Fig.~\ref{fig:learning} also shows that combinations of only the simple scalar-valued descriptors (2b, 2b+EAM, and 2b+$N$EAM) have limits in accuracy that are reached already with very little training data (at around 20--30 \% of the full training set). The learning curves of the more flexible combinations (2b+3b, 2b+3b+EAM, 2b+SOAP, 2b+cSOAP) are also quite different. For crystals, 2b+3b and 2b+3b+EAM both reach an acceptable 10 meV/atom test error for crystals already with 25--35 \% of the full training data, while 2b+SOAP needs the full training set and 2b+cSOAP 60--70 \% to reach this level of accuracy. With the full training set, 2b+3b, 2b+3b+EAM, and 2b+cSOAP reach 2--3 meV/atom test errors for crystals with trends that still display learning if more training data were to be added.

We can conclude from Fig.~\ref{fig:learning} that adding more data would mostly benefit the accuracy for energies of crystalline structures, and especially for 2b+SOAP. For liquids, all combinations of descriptors show a weak almost-converged level of accuracy, indicating that the error is limited by the flexibility of the descriptors and the finite interaction range, not the amount of training data. The faster convergence of the force errors compared to energies is likely because the actual training data is dominated by force components.

\begin{figure*}
    \centering
    \includegraphics[width=\linewidth]{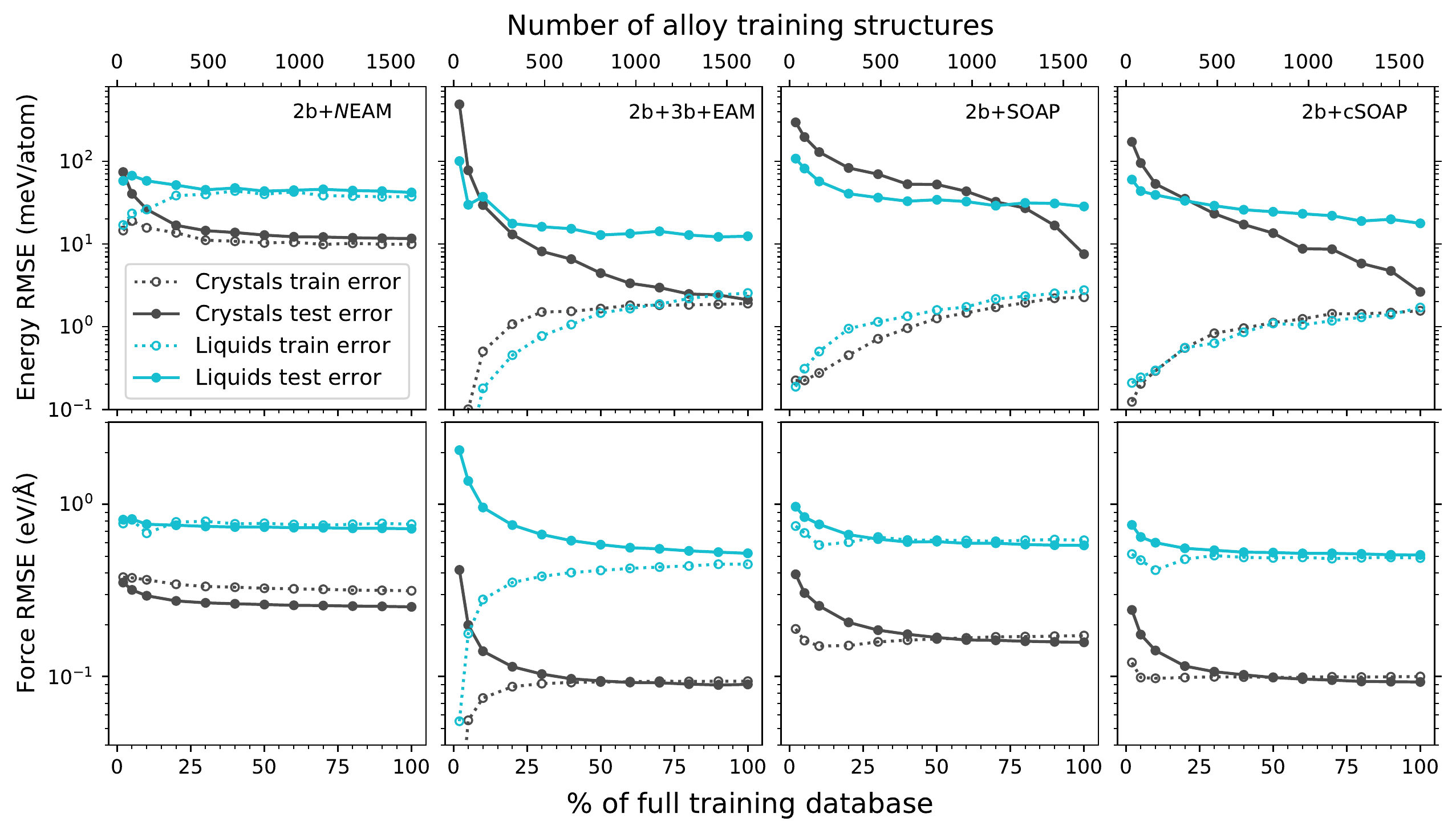}
    \caption{Learning curves of selected combinations of descriptors, showing the evolution of both train and test errors for the crystal and liquid alloy test sets.}
    \label{fig:learning2}
\end{figure*}

To estimate the theoretically achievable accuracies (given enough training data) in more detail, we plot in Fig.~\ref{fig:learning2} both the train and test errors for the four most interesting descriptor combinations. A good and well-trained model should produce similar train and test errors. As noted above, 2b+3b+EAM and 2b+cSOAP show the lowest test errors. 
Since the 2b+3b+EAM descriptors are relatively simple and low-dimensional, one could expect that given enough training data, the more flexible SOAP and cSOAP descriptors should eventually outperform all other potentials. However, Fig.~\ref{fig:learning2} suggests that at least for force components, 2b+SOAP and 2b+cSOAP are close to their achievable accuracy, as the learning rates are weak and the train and test errors have converged. For energies, extrapolation of the train- and test-error curves in Fig.~\ref{fig:learning2} indicates that 2b+SOAP and 2b+cSOAP would at best converge to similar or slightly better accuracy than 2b+3b+EAM. Given that the accuracy of 2b+3b+EAM for crystals is down to 2 meV/atom with the full training dataset, there is after all not much room for improvement.
Fig.~\ref{fig:learning2} also shows that 2b+$N$EAM quickly becomes oversaturated with data, showing almost no change in the train or test errors above 30 \% of the full training data.

From Fig.~\ref{fig:learning2} we can also conclude that the size of the full training dataset is appropriate for 2b+3b+EAM. With this set of descriptors, the train and test errors have converged towards each other in all cases except for the energies of liquids. The overfitting gaps between the train and test errors for the liquid energies indicate either a lack of data, an inherent limit in the flexibility of the descriptors with the given hyperparameters and cutoff range, or suboptimal regularisation parameters. Given the almost-flat learning curves for the test errors, adding more data would yield diminishing returns (and mainly raise the training errors towards the test errors). Tests also revealed that tuning the regularisation parameters had the same effect, i.e. raising the train errors without improving the test errors. Nevertheless, the liquid energy test error for 2b+3b+EAM potential reaches 12.4 meV/atom with the full training database, which we consider excellent accuracy for liquid five-element alloys.

\subsection{Speed versus accuracy}

\begin{figure}
    \centering
    \includegraphics[width=\linewidth]{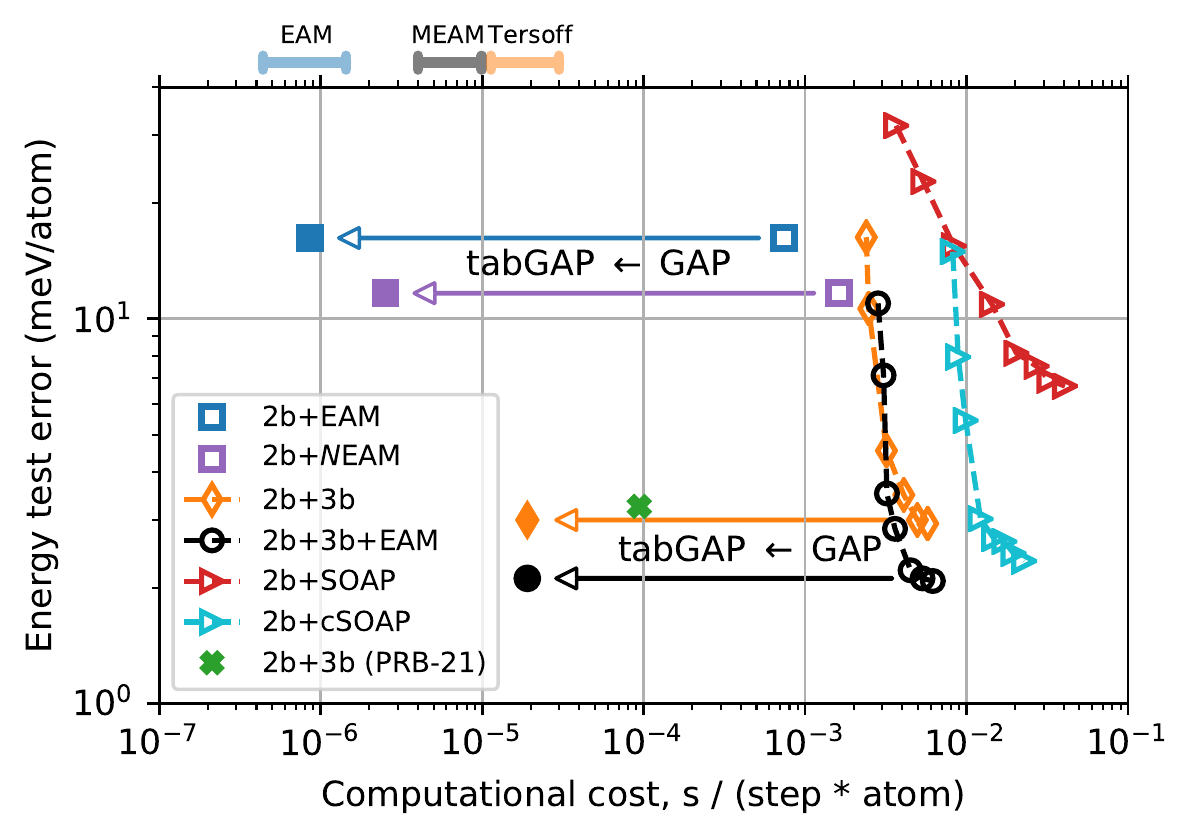}
    \caption{Accuracy versus computational cost of the GAPs with different combinations of descriptors and degrees of freedom (unfilled markers). The arrows show the speed-up following tabulation of the low-dimensional GAPs into the corresponding tabGAPs (filled markers). The accuracy is shown as the test errors for the crystal alloy test set. The previous 2b+3b tabGAP from Ref.~\cite{byggmastar_modeling_2021} is also shown. It is slower than the current 2b+3b version due to longer 3b cutoff. Approximate computational costs of some traditional interatomic potentials for bcc metals (EAM~\cite{daw_embedded-atom_1984}, MEAM~\cite{lee_second_2000}, Tersoff~\cite{tersoff_new_1988}) are shown above for comparison (from Ref.~\cite{byggmastar_multiscale_2022}).}
    \label{fig:speed}
\end{figure}

Perhaps of most practical relevance when comparing machine-learning potentials is the balance between speed and accuracy. Fig.~\ref{fig:speed} shows again the crystal test errors but as functions of the computational cost in molecular dynamics simulations. The cost is expressed in seconds per time step per atom, tested with a 1024-atom MoNbTaVW lattice on a single i5-7500 3.4 GHz CPU core with a \textsc{gnu} compilation of \textsc{lammps}. The different points for the same set of descriptors correspond to using different degrees of freedom (from Fig.~\ref{fig:nsparse}), which affects the computational speed significantly. Fig.~\ref{fig:speed} also shows the speedup and final computational cost of the tabGAPs, i.e. the tabulated versions of the 2b+EAM, 2b+$N$EAM, 2b+3b, and 2b+3b+EAM GAPs. Note that since all tabGAPs are evaluated using tabulated grids of energies, their speed is independent of the degrees of freedom used in the underlying GAP.

Fig.~\ref{fig:err_cfgs} also includes our previous 2b+3b tabGAP from~\cite{byggmastar_modeling_2021} for comparison, which is not exactly the same as the 2b+3b tabGAP trained here and shown in previous figures. The difference is in the cutoff distance of the 3b descriptor, which was 5 Å in~\cite{byggmastar_modeling_2021} and 4.1 Å in this work.

It is clear from Fig.~\ref{fig:speed} that the 2b+3b+EAM tabGAP is the most practically useful potential, being the most accurate while also 2--3 orders of magnitude faster than the 2b+SOAP and 2b+cSOAP GAPs. However, we should note that the tested cSOAP implementation is not fully optimised, and that we here only included the most accurate but computationally most costly compressed-SOAP version~\cite{darby_compressing_2021}. Ref.~\cite{darby_compressing_2021} discusses many more cSOAP options that are faster but not as accurate, although they are still at least two orders of magnitude slower than the 2b+3b+EAM tabGAP.

The 2b+$N$EAM tabGAP (or even 2b+EAM) is not nearly as accurate as 2b+3b+EAM but is significantly faster. In fact, a 2b+EAM tabGAP runs simply as a normal EAM potential using existing implementations. Hence, they can be considered extremely data-efficient and fast options when developing potentials for many-element alloys, but only if one can accept sacrificing meV/atom-accuracy in favour of computational speed.

\subsection{tabGAP for Mo--Nb--Ta--V--W high-entropy alloys}

\begin{figure}
    \centering
    \includegraphics[width=\linewidth]{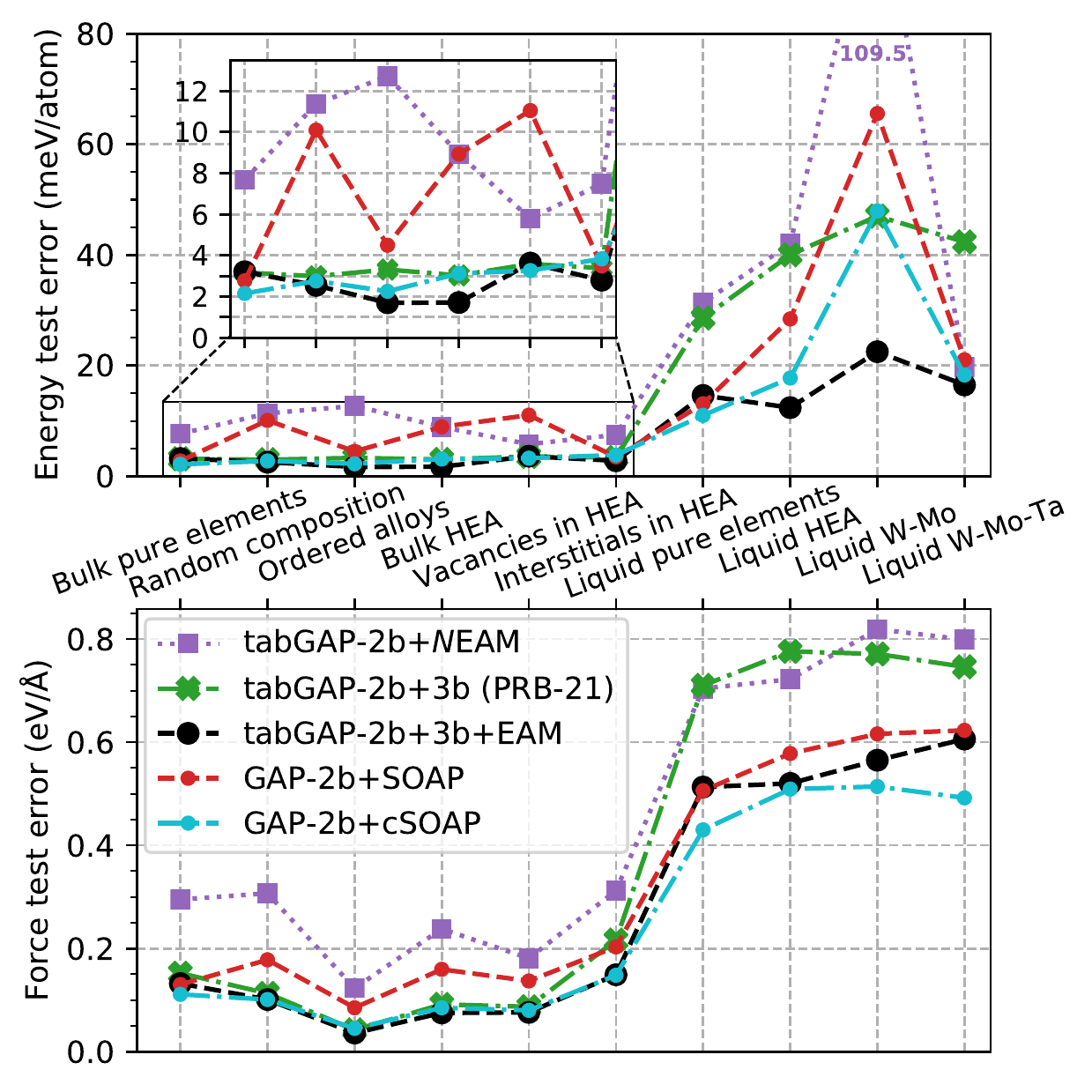}
    \caption{Test errors for the final tabGAPs and GAPs trained to the full Mo--Nb--Ta--V--W dataset, shown separately for the different classes of structures in the crystal and liquid test sets. HEA is short for the equiatomic MoNbTaVW high-entropy alloy composition. Test data for pure elements~\cite{byggmastar_gaussian_2020} and liquid W--Mo and W--Mo--Ta from Refs.~\cite{nikoulis_machine-learning_2021,kauppala_interatomic_2021} are also included (which were not part of the alloy test sets in previous figures). The 2b+3b tabGAP is the one from Ref.~\cite{byggmastar_modeling_2021}.}
    \label{fig:err_cfgs}
\end{figure}

Here, we report the final test errors in more detail for the most practically useful potentials trained to the full training dataset. In Fig.~\ref{fig:err_cfgs}, the test errors are separated by the different classes of structures in the test sets (where HEA refers to the MoNbTaVW high-entropy alloy). The figure also includes some pure-element test data from~\cite{byggmastar_gaussian_2020} as well as the liquids from the W--Mo dataset~\cite{nikoulis_machine-learning_2021} and the W--Mo--Ta dataset~\cite{kauppala_interatomic_2021}, which were not included in the alloy test sets in the previous figures. The 2b+3b tabGAP in Fig.~\ref{fig:err_cfgs} is the one from Ref~\cite{byggmastar_modeling_2021}.

Fig.~\ref{fig:err_cfgs} shows that the 2b+3b+EAM tabGAP is more accurate than the previous 2b+3b tabGAP~\cite{byggmastar_modeling_2021} for all classes of structures, with the biggest improvement for liquids. Adding the EAM descriptor is therefore beneficial for both speed and accuracy. First, it decreases the liquid test errors by factors of 2--3 and provides similar test errors for the crystal test errors. Second, it allows using a shorter 3b cutoff without losing accuracy, which makes the present 2b+3b+EAM tabGAP around 4 times faster than the previous 2b+3b tabGAP~\cite{byggmastar_modeling_2021}.

Fig.~\ref{fig:err_cfgs} shows that 2b+3b+EAM and 2b+cSOAP are similar in accuracy for most classes of materials, with the former overall more accurate for liquid energies but less accurate for liquid forces. The W--Mo liquids provide an interesting test set, as Fig.~\ref{fig:err_cfgs} shows that all potentials are significantly less accurate for their energies (although the same trend is not visible for the force test errors). The reason is that the W--Mo test data includes liquids spanning the entire composition range, while the HEA and W--Mo--Ta test data only contain liquids with the equiatomic composition. Hence, the W--Mo test set provides a good test of transferability, as the Mo--Nb--Ta--V--W training set also only contains equiatomic liquid compositions (but for all binary, ternary, quaternary, and quinary alloys). Fig.~\ref{fig:err_cfgs} shows that the 2b+3b+EAM tabGAP is most transferable to the W--Mo liquid compositions, while 2b+$N$EAM and 2b+SOAP show the poorest transferability.

It is important to remember that to validate a potential for molecular dynamics simulations, only looking at statically computed test errors is dangerous and does not ensure that the potential behaves physically well in simulations. It is crucial to perform tests of actual material properties and to run finite-temperature simulations. We used the previous 2b+3b tabGAP~\cite{byggmastar_modeling_2021} extensively in MD simulations and validated it against material properties computed with DFT. Since the present 2b+3b+EAM tabGAP is trained to the same data, we expect it to produce similar (but more accurate) material behaviour and lead to the same conclusions as in Ref.~\cite{byggmastar_modeling_2021} for the MoNbTaVW high-entropy alloy. We confirmed this by computing the lattice constant, mixing energy, bulk modulus, melting temperature, vacancy formation energies and relaxation volumes, as well as the short-range-order parameters in equiatomic MoNbTaVW optimised by hybrid Metropolis Monte Carlo and MD simulations at 300 K. The results are given in Table~\ref{tab:HEA} and calculated as the average of 50 different randomly ordered lattices (except for the melting temperature, which is simulated using the solid-liquid interface method and sampling different temperatures~\cite{Mor94}). The uncertainty of the DFT mixing energy and lattice constant are larger than for tabGAP, since for DFT we used small 54-atom boxes while for the tabGAPs we used 2000-atom boxes. The methods are explained in more detail in~\cite{byggmastar_modeling_2021}.

\begin{table*}
    \centering
    \caption{Properties of equiatomic MoNbTaVW predicted by the tabGAP 2b+3b+EAM compared to the previous tabGAP 2b+3b from Ref.~\cite{byggmastar_modeling_2021}, DFT from Ref.~\cite{byggmastar_modeling_2021}, and experimental estimates calculated as averages of the pure metals (Vegard's law) using data from Ref.~\cite{rumble_crc_2019}. $a$ is the lattice constant, $E_\mathrm{mix}$ is the mixing energy, $B$ is the bulk modulus, $E^\mathrm{f}_\mathrm{vac.}$ the average vacancy formation energy, $\Omega^\mathrm{rel.}_\mathrm{vac.}$ the vacancy relaxation volume, and $T_\mathrm{melt}$ the melting temperature. All tabGAP and DFT values except the melting temperature are the averages and standard deviation of 50 different randomly ordered lattices.}
    \begin{tabular}{lllll}
         \toprule
         & tabGAP 2b+3b+EAM & tabGAP 2b+3b~\cite{byggmastar_modeling_2021} & DFT~\cite{byggmastar_modeling_2021} & Expt. (Vegard's law) \\
         \midrule
         $a$ (Å) & $3.195 \pm 0.001$ & $3.195 \pm 0.001$ & $3.195 \pm 0.004$ & 3.188 \\
         $E_\mathrm{mix}$ (meV/atom) & $-38.3 \pm 0.7$ & $-41.8 \pm 0.7$ & $-43.8 \pm 7.3$  \\
         $B$ (GPa) & $216.4 \pm 0.1$ & $210.4 \pm 0.3$ & & 217.6 \\
         $E^\mathrm{f}_\mathrm{vac.}$ (eV) & $3.29 \pm 0.03$ & $3.13 \pm 0.04$ & $3.33 \pm 0.05$ \\
         $\Omega^\mathrm{rel.}_\mathrm{vac.}$ (at. vol.) & $-0.33 \pm 0.01$ & $-0.39 \pm 0.01$ & $-0.34 \pm 0.01$ \\
         $T_\mathrm{melt}$ (K) & $2760 \pm 20$ & $2760 \pm 20$ & & 2961 \\
         \bottomrule
    \end{tabular}
    \label{tab:HEA}
\end{table*}

Table~\ref{tab:HEA} shows that the present 2b+3b+EAM tabGAP predicts overall similar material properties to the previous 2b+3b tabGAP from Ref.~\cite{byggmastar_modeling_2021}, and in good agreement with the DFT results (also from \cite{byggmastar_modeling_2021}). Most noteworthy is the improvement in the vacancy properties, for which the 2b+3b+EAM tabGAP matches the DFT results well. For comparison, we also compare to approximate experimental values calculated with Vegard's law, i.e. the concentration-weighted mean of the pure-metal properties. The experimental Vegard's law produces similar values to the tabGAP predictions, with a 7 \% difference in the melting point.

The short-range order parameters are almost identical in the present tabGAP, showing local ordering of primarily the MoTa binary but also VW and MoNb. The short-range order parameters of the Mo--Ta, V--W, and Mo--Nb first-nearest-neighbour pairs are $-1.42$, $-0.60$, and $-0.80$ in the present 2b+3b+EAM tabGAP and $-1.37$, $-0.66$, and $-0.78$ in the previous 2b+3b tabGAP.

\begin{figure}
    \centering
    \includegraphics[width=\linewidth]{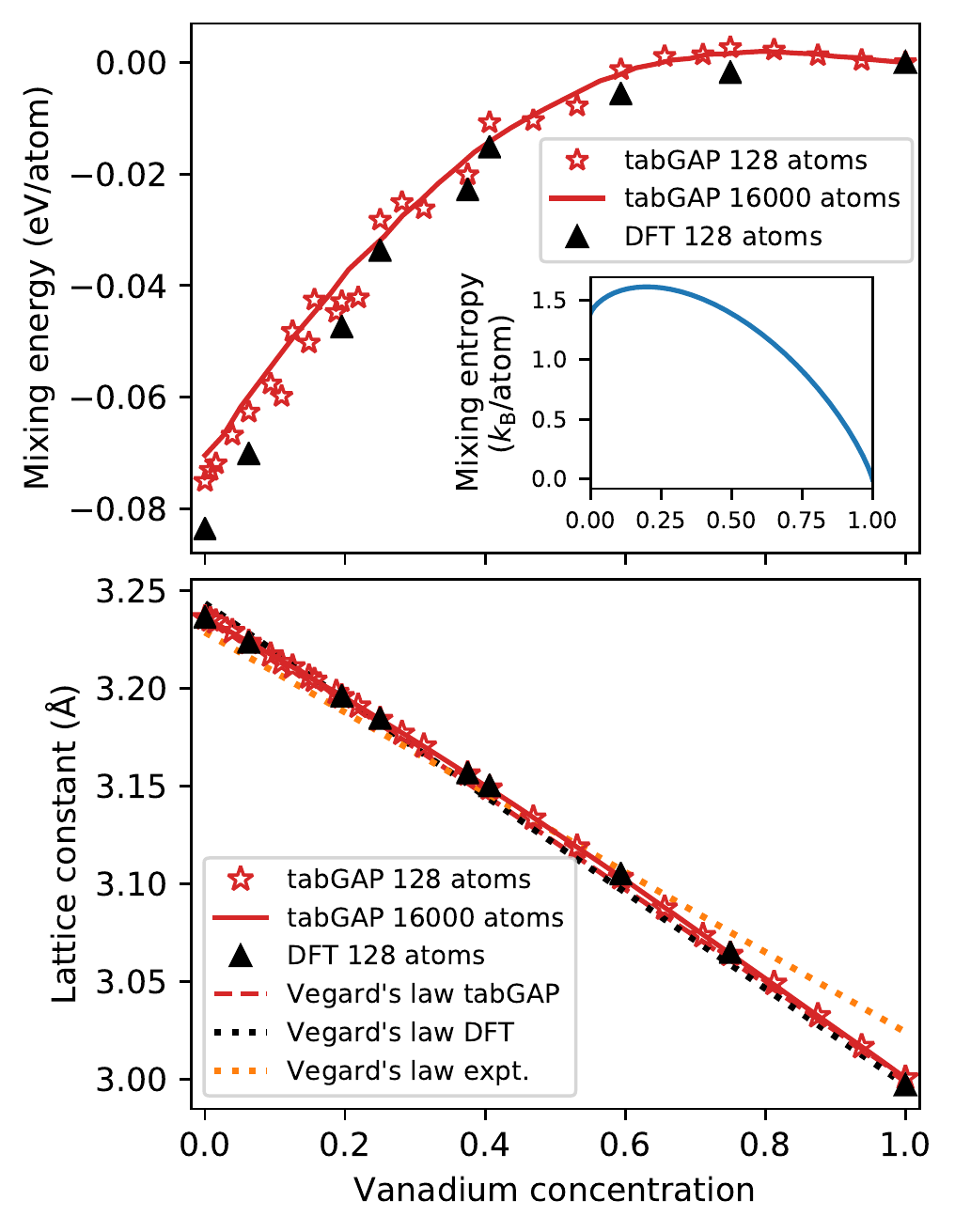}
    \caption{Mixing energy (top) and lattice constant (bottom) of randomly ordered Mo--Nb--Ta--V--W alloys with increasing vanadium concentration ($x$ in Mo$_{(1-x)/4}$Nb$_{(1-x)/4}$Ta$_{(1-x)/4}$V$_x$W$_{(1-x)/4}$). The results are compared between the tabGAP 2b+3b+EAM and DFT. For the lattice constant we also compare with Vegard's law, i.e. the weighted average of the pure-element lattice constants. The inset shows the analytically calculated  mixing entropy $\Delta S_\mathrm{mix}$ which reaches its maximum for the equiatomic MoNbTaVW composition $x=0.2$, for reference.}
    \label{fig:Vscan}
\end{figure}

As an additional test, we investigated the effect of vanadium concentration on the lattice constant and mixing energy in both DFT and with the 2b+3b+EAM tabGAP. For DFT, we used \textsc{vasp}~\cite{kresse_ab_1993} at the GGA-PBE~\cite{perdew_generalized_1996} level with the same input parameters as the training and testing data (see Ref.~\cite{byggmastar_modeling_2021}). We fully relaxed randomly ordered alloys with increasing vanadium concentration from the vanadium-free equiatomic alloy to pure vanadium, i.e. Mo$_{(1-x)/4}$Nb$_{(1-x)/4}$Ta$_{(1-x)/4}$V$_x$W$_{(1-x)/4}$ alloys with $x$ varying from 0 to 1. From the relaxed alloys, we compute the mixing energies and the lattice constants. For DFT, we used 128-atom boxes, and for tabGAP both 128-atom boxes but also larger boxes of 16 000 atoms to obtain a smoother and more reliable trend. The results are shown in Fig.~\ref{fig:Vscan}. For the lattice constant we also plot the trend given by Vegard's law. The tabGAP reproduces the DFT data well. For the mixing energy, there is a small systematic offset of a few meV/atom, which is the expected accuracy of the tabGAP. In Fig.~\ref{fig:Vscan} we also plot the mixing entropy, $\Delta S_\mathrm{mix} = -Nk_\mathrm{B} \sum_i c_i \ln c_i$ where $c_i$ is the concentration of element $i$, to highlight the maximised mixing entropy for the equiatomic MoNbTaVW alloy.

We conclude that the new 2b+3b+EAM tabGAP will be useful in simulations of Mo--Nb--Ta--V--W alloys. It should be preferred over the previous version~\cite{byggmastar_modeling_2021}, given that it is both more accurate and 4 times faster.

\subsection{From pure elements to alloys}

\begin{figure}
    \centering
    \includegraphics[width=\linewidth]{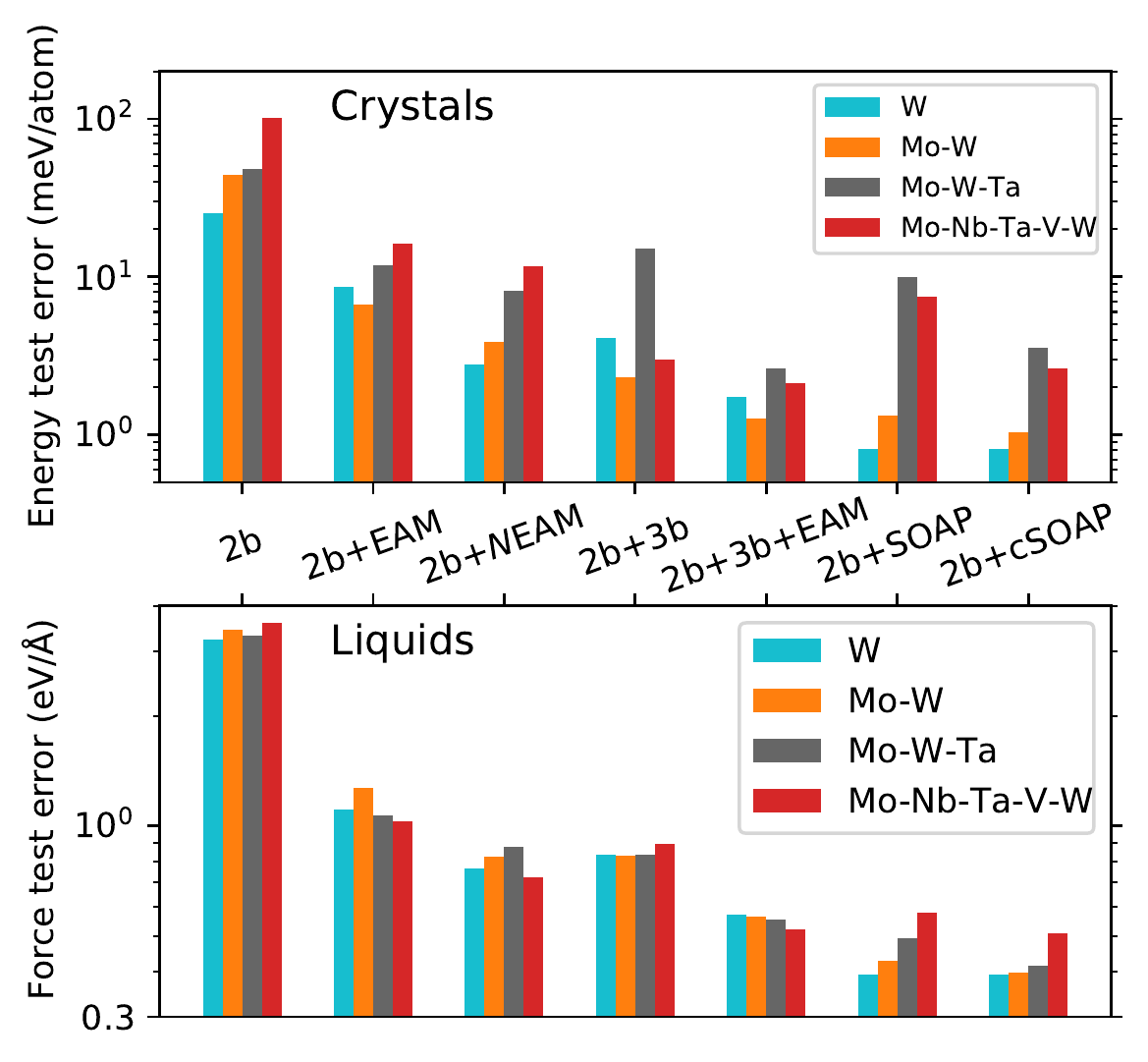}
    \caption{Test errors of potentials trained only to the pure-W dataset~\cite{byggmastar_machine-learning_2019}, to the W--Mo dataset~\cite{nikoulis_machine-learning_2021}, to the W--Mo--Ta dataset~\cite{kauppala_interatomic_2021}, and to the full Mo--Nb--Ta--V-W dataset.}
    \label{fig:from}
\end{figure}

The main aim of this work has been to study the learning curves and data efficiency of ML potentials for multicomponent alloys. As a final note, it is worth highlighting the differences in accuracy when going from pure-element potentials to the five-element Mo--Nb--Ta--V--W potentials. Fig.~\ref{fig:from} shows test errors from the previous figures for the Mo--Nb--Ta--V--W potentials, now compared to sets of potentials with the same combinations of descriptors but trained separately to pure W (using the data from~\cite{byggmastar_machine-learning_2019}), Mo--W binary alloys (using data from~\cite{nikoulis_machine-learning_2021}), and ternary Mo--W--Ta alloys (data from~\cite{kauppala_interatomic_2021}). We only show the force errors for liquids, which are more representative of the accuracy due to the limited number of liquid structures in the test sets.

Fig.~\ref{fig:from} shows that for pure W, the 2b+SOAP GAP is by far the most accurate potential with test errors below 1 meV/atom, which justifies its higher computational cost and leads to very accurate material properties~\cite{byggmastar_machine-learning_2019}. For the binary W--Mo potentials, the SOAP (and cSOAP) GAPs are still most accurate. However, for the ternary Mo--W--Ta potentials, the 2b+3b+EAM outperforms 2b+SOAP and matches 2b+cSOAP. The energy test errors for crystals are quite high for the Mo--W--Ta potentials, which is because the size of the dataset is relatively small. The accuracy of 2b+3b+EAM potentials is roughly the same regardless of number of elements in the training data, at around 2--3 meV/atom for crystals. In contrast, 2b+SOAP (and to a lesser extent 2b+cSOAP) show clear trends of increased errors as more elements are added. This is particularly clear in the liquid force errors for 2b+SOAP in Fig.~\ref{fig:from}. It is likely a consequence of the dimensionality of the SOAP vector scaling quadratically with number of elements, so that the descriptor space with five elements is unnecessarily high-dimensional. The 2b+cSOAP dimensionality scales linearly with number of elements~\cite{darby_compressing_2021}, and hence provide increasingly bigger improvements over 2b+SOAP as more elements are added.

\section{Conclusions}

We have systematically investigated the data efficiency, achievable accuracy, and computational cost of machine-learning interatomic potentials for many-element alloys using different combinations of atomistic descriptors. The main conclusion is that when training ML potentials for multicomponent alloys, one can achieve remarkably good accuracy with simple low-dimensional descriptors for the local atomic environments. This is an encouraging and practically helpful result, as we showed that low-dimensional ML potentials are also quite data-efficient, reaching 2 meV/atom accuracy for five-element alloys with a modest amount of training data. Additionally, low-dimensional ML potentials are computationally efficient compared to ML potentials using many-body descriptors since they can be tabulated and evaluated using conventional interpolation methods.

Our results show that when developing ML interatomic potentials for many-element systems, it is not necessary to use flexible and high-dimensional many-body descriptors (such as SOAP). Instead, low-dimensional descriptors can provide the desired near-quantum accuracy, with the added benefits of making the potential more interpretable, data-efficient, faster, and also seemingly more transferable than high-dimensional ML potentials. However, we also showed that for single-element materials where data efficiency is a lesser issue, the accuracy of high-dimensional ML potentials (such as SOAP-GAP) is still out of reach for low-dimensional ML potentials (such as tabGAP).

We also introduced a simple scalar multi-element embedded atom method density descriptor, and showed that a GAP using two-body, three-body, and the EAM descriptor performs best among the tested combinations of descriptors for the five-element Mo--Nb--Ta--V--W system. When tabulated into a tabGAP, the computational cost is reduced by two orders of magnitude, making molecular dynamics simulations with millions of atoms and nanosecond time scales well within reach. The tabGAP is available from Ref.~\cite{byggmastar_ida_data_2022} and is, because of the EAM descriptor, an improvement over our previously published tabGAP~\cite{byggmastar_modeling_2021} in both speed (around 4 times faster) and accuracy (especially for the liquid phase).

\section*{Acknowledgements}

This work was partially supported by the Academy of Finland Flagship programme: Finnish Center for Artificial Intelligence FCAI.
This work has partially been carried out within the framework of the EUROfusion Consortium, funded by the European Union via the Euratom Research and Training Programme (Grant Agreement No 101052200 — EUROfusion). Views and opinions expressed are however those of the author(s) only and do not necessarily reflect those of the European Union or the European Commission. Neither the European Union nor the European Commission can be held responsible for them.
Grants of computer capacity from CSC - IT Center for Science are gratefully acknowledged.

\appendix
\section{Pair-density functions}
\label{sec:appendix}

\begin{figure}
    \centering
    \includegraphics[width=\linewidth]{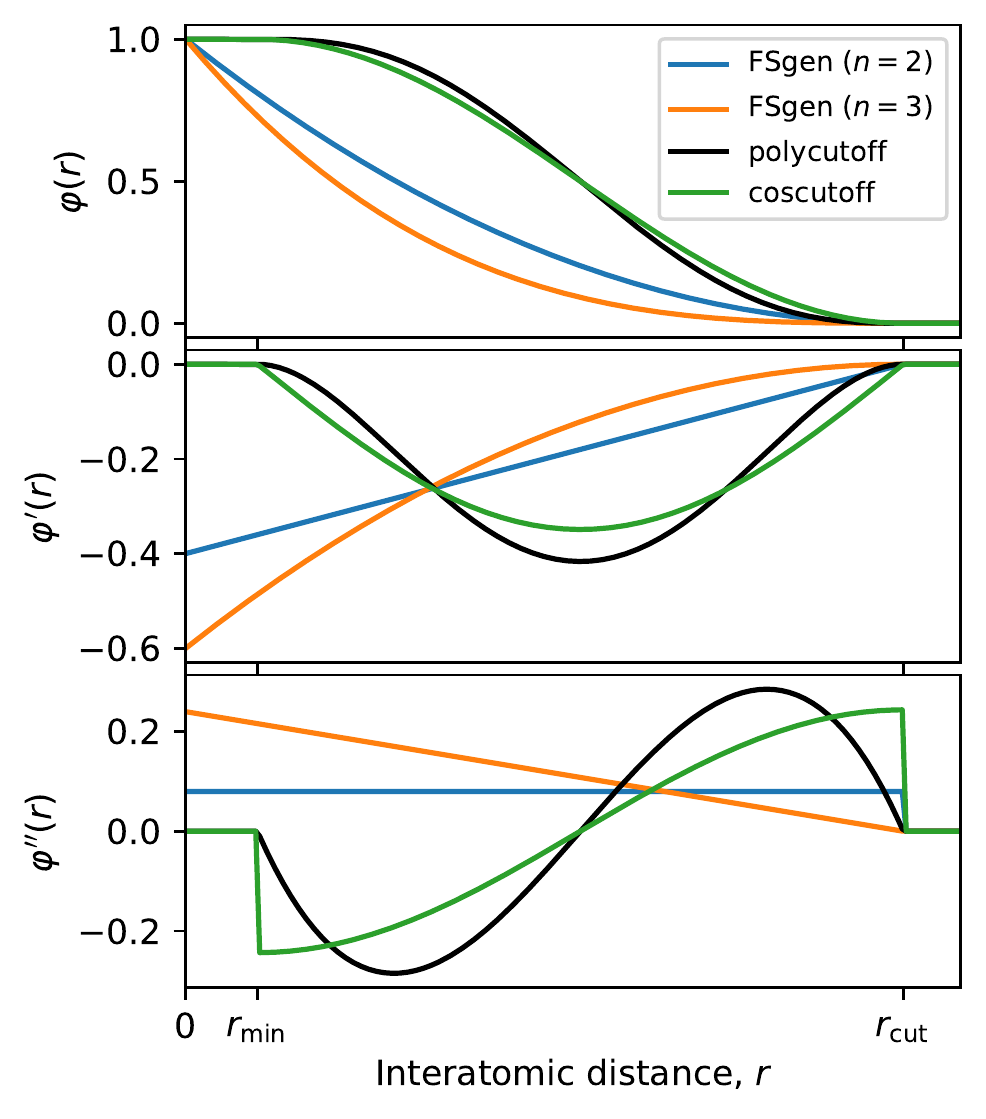}
    \caption{Examples of the different pair-density functions $\varphi (r)$ and their first and second derivatives.}
    \label{fig:varphi}
\end{figure}

The following are the pair-density functions implemented into \textsc{quip}~\cite{QUIP} for the EAM descriptor discussed in the main text:
\begin{equation}
   \varphi(r) =
   \begin{cases}
    \displaystyle \left(1 - \frac{r}{r_\mathrm{cut}}\right)^n, ~r < r_\mathrm{cut} \\
    0, ~r \geq r_\mathrm{cut}, \\
   \end{cases}
   \label{eq:pairdens1}
\end{equation}
\begin{equation}
   \varphi(r) =
   \begin{cases}
    1, ~r \leq r_\mathrm{min} \\
    \displaystyle 1 - \chi^3 (6 \chi^2 - 15 \chi + 10), ~r_\mathrm{min} < r < r_\mathrm{cut} \\
    0, ~r \geq r_\mathrm{cut}, \\
   \end{cases}
   \label{eq:pairdens2}
\end{equation}
\begin{equation}
   \varphi(r) =
   \begin{cases}
    1, ~r \leq r_\mathrm{min} \\
    \displaystyle \frac{1}{2} \left[1 + \cos (\pi \chi)\right], ~r_\mathrm{min} < r < r_\mathrm{cut} \\
    0, ~r \geq r_\mathrm{cut}, \\
   \end{cases}
   \label{eq:pairdens3}
\end{equation}
where $\chi = (r - r_\mathrm{min})/(r_\mathrm{cut} - r_\mathrm{min})$. The first function can be considered a generalised Finnis-Sinclair-like~\cite{finnis_simple_1984} pair-density function (hence named \texttt{FSgen} in the code) and the other two are common ``cutoff-like'' functions (\texttt{polycutoff} and \texttt{coscutoff} in the code). \texttt{coscutoff} is the cutoff function used in Tersoff-like potentials~\cite{tersoff_new_1988} and also for most descriptors implemented in the GAP code. The user-defined parameters are the polynomial order $n$ (integer $\geq 2$), the cutoff radius $r_\mathrm{cut}$, and $r_\mathrm{min}$ ($0 \leq r_\mathrm{min} < r_\mathrm{cut}$). The functions are illustrated in Fig.~\ref{fig:varphi} along with the first and second derivatives. \texttt{FSgen} with $n \geq 3$ and the \texttt{polycutoff} functions have continuous second derivatives, but not \texttt{coscutoff}.

\bibliography{mybib}

\begin{thebibliography}{50}%
\makeatletter
\providecommand \@ifxundefined [1]{%
 \@ifx{#1\undefined}
}%
\providecommand \@ifnum [1]{%
 \ifnum #1\expandafter \@firstoftwo
 \else \expandafter \@secondoftwo
 \fi
}%
\providecommand \@ifx [1]{%
 \ifx #1\expandafter \@firstoftwo
 \else \expandafter \@secondoftwo
 \fi
}%
\providecommand \natexlab [1]{#1}%
\providecommand \enquote  [1]{``#1''}%
\providecommand \bibnamefont  [1]{#1}%
\providecommand \bibfnamefont [1]{#1}%
\providecommand \citenamefont [1]{#1}%
\providecommand \href@noop [0]{\@secondoftwo}%
\providecommand \href [0]{\begingroup \@sanitize@url \@href}%
\providecommand \@href[1]{\@@startlink{#1}\@@href}%
\providecommand \@@href[1]{\endgroup#1\@@endlink}%
\providecommand \@sanitize@url [0]{\catcode `\\12\catcode `\$12\catcode
  `\&12\catcode `\#12\catcode `\^12\catcode `\_12\catcode `\%12\relax}%
\providecommand \@@startlink[1]{}%
\providecommand \@@endlink[0]{}%
\providecommand \url  [0]{\begingroup\@sanitize@url \@url }%
\providecommand \@url [1]{\endgroup\@href {#1}{\urlprefix }}%
\providecommand \urlprefix  [0]{URL }%
\providecommand \Eprint [0]{\href }%
\providecommand \doibase [0]{https://doi.org/}%
\providecommand \selectlanguage [0]{\@gobble}%
\providecommand \bibinfo  [0]{\@secondoftwo}%
\providecommand \bibfield  [0]{\@secondoftwo}%
\providecommand \translation [1]{[#1]}%
\providecommand \BibitemOpen [0]{}%
\providecommand \bibitemStop [0]{}%
\providecommand \bibitemNoStop [0]{.\EOS\space}%
\providecommand \EOS [0]{\spacefactor3000\relax}%
\providecommand \BibitemShut  [1]{\csname bibitem#1\endcsname}%
\let\auto@bib@innerbib\@empty
\bibitem [{\citenamefont {Behler}(2016)}]{behler_perspective_2016}%
  \BibitemOpen
  \bibfield  {author} {\bibinfo {author} {\bibfnamefont {J.}~\bibnamefont
  {Behler}},\ }\bibfield  {title} {\bibinfo {title} {Perspective: {{Machine}}
  learning potentials for atomistic simulations},\ }\href
  {https://doi.org/10.1063/1.4966192} {\bibfield  {journal} {\bibinfo
  {journal} {J. Chem. Phys.}\ }\textbf {\bibinfo {volume} {145}},\ \bibinfo
  {pages} {170901} (\bibinfo {year} {2016})}\BibitemShut {NoStop}%
\bibitem [{\citenamefont {Deringer}\ \emph {et~al.}(2021)\citenamefont
  {Deringer}, \citenamefont {Bart{\'o}k}, \citenamefont {Bernstein},
  \citenamefont {Wilkins}, \citenamefont {Ceriotti},\ and\ \citenamefont
  {Cs{\'a}nyi}}]{deringer_gaussian_2021}%
  \BibitemOpen
  \bibfield  {author} {\bibinfo {author} {\bibfnamefont {V.~L.}\ \bibnamefont
  {Deringer}}, \bibinfo {author} {\bibfnamefont {A.~P.}\ \bibnamefont
  {Bart{\'o}k}}, \bibinfo {author} {\bibfnamefont {N.}~\bibnamefont
  {Bernstein}}, \bibinfo {author} {\bibfnamefont {D.~M.}\ \bibnamefont
  {Wilkins}}, \bibinfo {author} {\bibfnamefont {M.}~\bibnamefont {Ceriotti}},\
  and\ \bibinfo {author} {\bibfnamefont {G.}~\bibnamefont {Cs{\'a}nyi}},\
  }\bibfield  {title} {\bibinfo {title} {Gaussian {{Process Regression}} for
  {{Materials}} and {{Molecules}}},\ }\href
  {https://doi.org/10.1021/acs.chemrev.1c00022} {\bibfield  {journal} {\bibinfo
   {journal} {Chem. Rev.}\ }\textbf {\bibinfo {volume} {121}},\ \bibinfo
  {pages} {10073} (\bibinfo {year} {2021})}\BibitemShut {NoStop}%
\bibitem [{\citenamefont {{Y. Mishin}}(2021)}]{Mis21}%
  \BibitemOpen
  \bibfield  {author} {\bibinfo {author} {\bibnamefont {{Y. Mishin}}},\
  }\bibfield  {title} {\bibinfo {title} {{Machine-learning interatomic
  potentials for materials science}},\ }\href
  {https://doi.org/10.1016/j.actamat.2021.116980} {\bibfield  {journal}
  {\bibinfo  {journal} {Acta Materialia}\ }\textbf {\bibinfo {volume} {214}},\
  \bibinfo {pages} {116980} (\bibinfo {year} {2021})}\BibitemShut {NoStop}%
\bibitem [{\citenamefont {Behler}\ and\ \citenamefont
  {Parrinello}(2007)}]{behler_generalized_2007}%
  \BibitemOpen
  \bibfield  {author} {\bibinfo {author} {\bibfnamefont {J.}~\bibnamefont
  {Behler}}\ and\ \bibinfo {author} {\bibfnamefont {M.}~\bibnamefont
  {Parrinello}},\ }\bibfield  {title} {\bibinfo {title} {Generalized
  {{Neural}}-{{Network Representation}} of {{High}}-{{Dimensional
  Potential}}-{{Energy Surfaces}}},\ }\bibfield  {journal} {\bibinfo  {journal}
  {Physical Review Letters}\ }\textbf {\bibinfo {volume} {98}},\ \href
  {https://doi.org/10.1103/PhysRevLett.98.146401}
  {10.1103/PhysRevLett.98.146401} (\bibinfo {year} {2007})\BibitemShut
  {NoStop}%
\bibitem [{\citenamefont {Bart{\'o}k}\ \emph {et~al.}(2010)\citenamefont
  {Bart{\'o}k}, \citenamefont {Payne}, \citenamefont {Kondor},\ and\
  \citenamefont {Cs{\'a}nyi}}]{bartok_gaussian_2010}%
  \BibitemOpen
  \bibfield  {author} {\bibinfo {author} {\bibfnamefont {A.~P.}\ \bibnamefont
  {Bart{\'o}k}}, \bibinfo {author} {\bibfnamefont {M.~C.}\ \bibnamefont
  {Payne}}, \bibinfo {author} {\bibfnamefont {R.}~\bibnamefont {Kondor}},\ and\
  \bibinfo {author} {\bibfnamefont {G.}~\bibnamefont {Cs{\'a}nyi}},\ }\bibfield
   {title} {\bibinfo {title} {Gaussian {{Approximation Potentials}}: {{The
  Accuracy}} of {{Quantum Mechanics}}, without the {{Electrons}}},\ }\bibfield
  {journal} {\bibinfo  {journal} {Physical Review Letters}\ }\textbf {\bibinfo
  {volume} {104}},\ \href {https://doi.org/10.1103/PhysRevLett.104.136403}
  {10.1103/PhysRevLett.104.136403} (\bibinfo {year} {2010})\BibitemShut
  {NoStop}%
\bibitem [{\citenamefont {Thompson}\ \emph {et~al.}(2015)\citenamefont
  {Thompson}, \citenamefont {Swiler}, \citenamefont {Trott}, \citenamefont
  {Foiles},\ and\ \citenamefont {Tucker}}]{thompson_spectral_2015}%
  \BibitemOpen
  \bibfield  {author} {\bibinfo {author} {\bibfnamefont {A.~P.}\ \bibnamefont
  {Thompson}}, \bibinfo {author} {\bibfnamefont {L.~P.}\ \bibnamefont
  {Swiler}}, \bibinfo {author} {\bibfnamefont {C.~R.}\ \bibnamefont {Trott}},
  \bibinfo {author} {\bibfnamefont {S.~M.}\ \bibnamefont {Foiles}},\ and\
  \bibinfo {author} {\bibfnamefont {G.~J.}\ \bibnamefont {Tucker}},\ }\bibfield
   {title} {\bibinfo {title} {Spectral neighbor analysis method for automated
  generation of quantum-accurate interatomic potentials},\ }\href
  {https://doi.org/10.1016/j.jcp.2014.12.018} {\bibfield  {journal} {\bibinfo
  {journal} {Journal of Computational Physics}\ }\textbf {\bibinfo {volume}
  {285}},\ \bibinfo {pages} {316} (\bibinfo {year} {2015})}\BibitemShut
  {NoStop}%
\bibitem [{\citenamefont {Drautz}(2019)}]{drautz_atomic_2019}%
  \BibitemOpen
  \bibfield  {author} {\bibinfo {author} {\bibfnamefont {R.}~\bibnamefont
  {Drautz}},\ }\bibfield  {title} {\bibinfo {title} {Atomic cluster expansion
  for accurate and transferable interatomic potentials},\ }\bibfield  {journal}
  {\bibinfo  {journal} {Physical Review B}\ }\textbf {\bibinfo {volume} {99}},\
  \href {https://doi.org/10.1103/PhysRevB.99.014104}
  {10.1103/PhysRevB.99.014104} (\bibinfo {year} {2019})\BibitemShut {NoStop}%
\bibitem [{\citenamefont {Fan}\ \emph {et~al.}(2021)\citenamefont {Fan},
  \citenamefont {Zeng}, \citenamefont {Zhang}, \citenamefont {Wang},
  \citenamefont {Song}, \citenamefont {Dong}, \citenamefont {Chen},\ and\
  \citenamefont {{Ala-Nissila}}}]{fan_neuroevolution_2021}%
  \BibitemOpen
  \bibfield  {author} {\bibinfo {author} {\bibfnamefont {Z.}~\bibnamefont
  {Fan}}, \bibinfo {author} {\bibfnamefont {Z.}~\bibnamefont {Zeng}}, \bibinfo
  {author} {\bibfnamefont {C.}~\bibnamefont {Zhang}}, \bibinfo {author}
  {\bibfnamefont {Y.}~\bibnamefont {Wang}}, \bibinfo {author} {\bibfnamefont
  {K.}~\bibnamefont {Song}}, \bibinfo {author} {\bibfnamefont {H.}~\bibnamefont
  {Dong}}, \bibinfo {author} {\bibfnamefont {Y.}~\bibnamefont {Chen}},\ and\
  \bibinfo {author} {\bibfnamefont {T.}~\bibnamefont {{Ala-Nissila}}},\
  }\bibfield  {title} {\bibinfo {title} {Neuroevolution machine learning
  potentials: {{Combining}} high accuracy and low cost in atomistic simulations
  and application to heat transport},\ }\href
  {https://doi.org/10.1103/PhysRevB.104.104309} {\bibfield  {journal} {\bibinfo
   {journal} {Phys. Rev. B}\ }\textbf {\bibinfo {volume} {104}},\ \bibinfo
  {pages} {104309} (\bibinfo {year} {2021})}\BibitemShut {NoStop}%
\bibitem [{\citenamefont {Byggm{\"a}star}\ \emph {et~al.}(2021)\citenamefont
  {Byggm{\"a}star}, \citenamefont {Nordlund},\ and\ \citenamefont
  {Djurabekova}}]{byggmastar_modeling_2021}%
  \BibitemOpen
  \bibfield  {author} {\bibinfo {author} {\bibfnamefont {J.}~\bibnamefont
  {Byggm{\"a}star}}, \bibinfo {author} {\bibfnamefont {K.}~\bibnamefont
  {Nordlund}},\ and\ \bibinfo {author} {\bibfnamefont {F.}~\bibnamefont
  {Djurabekova}},\ }\bibfield  {title} {\bibinfo {title} {Modeling refractory
  high-entropy alloys with efficient machine-learned interatomic potentials:
  {{Defects}} and segregation},\ }\href
  {https://doi.org/10.1103/PhysRevB.104.104101} {\bibfield  {journal} {\bibinfo
   {journal} {Phys. Rev. B}\ }\textbf {\bibinfo {volume} {104}},\ \bibinfo
  {pages} {104101} (\bibinfo {year} {2021})}\BibitemShut {NoStop}%
\bibitem [{\citenamefont {Li}\ \emph {et~al.}(2020)\citenamefont {Li},
  \citenamefont {Chen}, \citenamefont {Zheng}, \citenamefont {Zuo},\ and\
  \citenamefont {Ong}}]{li_complex_2020}%
  \BibitemOpen
  \bibfield  {author} {\bibinfo {author} {\bibfnamefont {X.-G.}\ \bibnamefont
  {Li}}, \bibinfo {author} {\bibfnamefont {C.}~\bibnamefont {Chen}}, \bibinfo
  {author} {\bibfnamefont {H.}~\bibnamefont {Zheng}}, \bibinfo {author}
  {\bibfnamefont {Y.}~\bibnamefont {Zuo}},\ and\ \bibinfo {author}
  {\bibfnamefont {S.~P.}\ \bibnamefont {Ong}},\ }\bibfield  {title} {\bibinfo
  {title} {Complex strengthening mechanisms in the {{NbMoTaW}} multi-principal
  element alloy},\ }\href {https://doi.org/10.1038/s41524-020-0339-0}
  {\bibfield  {journal} {\bibinfo  {journal} {npj Computational Materials}\
  }\textbf {\bibinfo {volume} {6}},\ \bibinfo {pages} {1} (\bibinfo {year}
  {2020})}\BibitemShut {NoStop}%
\bibitem [{\citenamefont {Hajinazar}\ \emph {et~al.}(2017)\citenamefont
  {Hajinazar}, \citenamefont {Shao},\ and\ \citenamefont
  {Kolmogorov}}]{hajinazar_stratified_2017-1}%
  \BibitemOpen
  \bibfield  {author} {\bibinfo {author} {\bibfnamefont {S.}~\bibnamefont
  {Hajinazar}}, \bibinfo {author} {\bibfnamefont {J.}~\bibnamefont {Shao}},\
  and\ \bibinfo {author} {\bibfnamefont {A.~N.}\ \bibnamefont {Kolmogorov}},\
  }\bibfield  {title} {\bibinfo {title} {Stratified construction of neural
  network based interatomic models for multicomponent materials},\ }\bibfield
  {journal} {\bibinfo  {journal} {Physical Review B}\ }\textbf {\bibinfo
  {volume} {95}},\ \href {https://doi.org/10.1103/PhysRevB.95.014114}
  {10.1103/PhysRevB.95.014114} (\bibinfo {year} {2017})\BibitemShut {NoStop}%
\bibitem [{\citenamefont {Rosenbrock}\ \emph {et~al.}(2021)\citenamefont
  {Rosenbrock}, \citenamefont {Gubaev}, \citenamefont {Shapeev}, \citenamefont
  {P{\'a}rtay}, \citenamefont {Bernstein}, \citenamefont {Cs{\'a}nyi},\ and\
  \citenamefont {Hart}}]{rosenbrock_machine-learned_2021}%
  \BibitemOpen
  \bibfield  {author} {\bibinfo {author} {\bibfnamefont {C.~W.}\ \bibnamefont
  {Rosenbrock}}, \bibinfo {author} {\bibfnamefont {K.}~\bibnamefont {Gubaev}},
  \bibinfo {author} {\bibfnamefont {A.~V.}\ \bibnamefont {Shapeev}}, \bibinfo
  {author} {\bibfnamefont {L.~B.}\ \bibnamefont {P{\'a}rtay}}, \bibinfo
  {author} {\bibfnamefont {N.}~\bibnamefont {Bernstein}}, \bibinfo {author}
  {\bibfnamefont {G.}~\bibnamefont {Cs{\'a}nyi}},\ and\ \bibinfo {author}
  {\bibfnamefont {G.~L.~W.}\ \bibnamefont {Hart}},\ }\bibfield  {title}
  {\bibinfo {title} {Machine-learned interatomic potentials for alloys and
  alloy phase diagrams},\ }\href {https://doi.org/10.1038/s41524-020-00477-2}
  {\bibfield  {journal} {\bibinfo  {journal} {npj Comput Mater}\ }\textbf
  {\bibinfo {volume} {7}},\ \bibinfo {pages} {1} (\bibinfo {year}
  {2021})}\BibitemShut {NoStop}%
\bibitem [{\citenamefont {Nikoulis}\ \emph {et~al.}(2021)\citenamefont
  {Nikoulis}, \citenamefont {Byggm{\"a}star}, \citenamefont {Kioseoglou},
  \citenamefont {Nordlund},\ and\ \citenamefont
  {Djurabekova}}]{nikoulis_machine-learning_2021}%
  \BibitemOpen
  \bibfield  {author} {\bibinfo {author} {\bibfnamefont {G.}~\bibnamefont
  {Nikoulis}}, \bibinfo {author} {\bibfnamefont {J.}~\bibnamefont
  {Byggm{\"a}star}}, \bibinfo {author} {\bibfnamefont {J.}~\bibnamefont
  {Kioseoglou}}, \bibinfo {author} {\bibfnamefont {K.}~\bibnamefont
  {Nordlund}},\ and\ \bibinfo {author} {\bibfnamefont {F.}~\bibnamefont
  {Djurabekova}},\ }\bibfield  {title} {\bibinfo {title} {Machine-learning
  interatomic potential for {{W}}\textendash{{Mo}} alloys},\ }\href
  {https://doi.org/10.1088/1361-648X/ac03d1} {\bibfield  {journal} {\bibinfo
  {journal} {J. Phys.: Condens. Matter}\ }\textbf {\bibinfo {volume} {33}},\
  \bibinfo {pages} {315403} (\bibinfo {year} {2021})}\BibitemShut {NoStop}%
\bibitem [{\citenamefont {Artrith}\ \emph {et~al.}(2017)\citenamefont
  {Artrith}, \citenamefont {Urban},\ and\ \citenamefont
  {Ceder}}]{artrith_efficient_2017}%
  \BibitemOpen
  \bibfield  {author} {\bibinfo {author} {\bibfnamefont {N.}~\bibnamefont
  {Artrith}}, \bibinfo {author} {\bibfnamefont {A.}~\bibnamefont {Urban}},\
  and\ \bibinfo {author} {\bibfnamefont {G.}~\bibnamefont {Ceder}},\ }\bibfield
   {title} {\bibinfo {title} {Efficient and accurate machine-learning
  interpolation of atomic energies in compositions with many species},\
  }\bibfield  {journal} {\bibinfo  {journal} {Physical Review B}\ }\textbf
  {\bibinfo {volume} {96}},\ \href {https://doi.org/10.1103/PhysRevB.96.014112}
  {10.1103/PhysRevB.96.014112} (\bibinfo {year} {2017})\BibitemShut {NoStop}%
\bibitem [{\citenamefont {Imbalzano}\ \emph {et~al.}(2018)\citenamefont
  {Imbalzano}, \citenamefont {Anelli}, \citenamefont {Giofr{\'e}},
  \citenamefont {Klees}, \citenamefont {Behler},\ and\ \citenamefont
  {Ceriotti}}]{imbalzano_automatic_2018}%
  \BibitemOpen
  \bibfield  {author} {\bibinfo {author} {\bibfnamefont {G.}~\bibnamefont
  {Imbalzano}}, \bibinfo {author} {\bibfnamefont {A.}~\bibnamefont {Anelli}},
  \bibinfo {author} {\bibfnamefont {D.}~\bibnamefont {Giofr{\'e}}}, \bibinfo
  {author} {\bibfnamefont {S.}~\bibnamefont {Klees}}, \bibinfo {author}
  {\bibfnamefont {J.}~\bibnamefont {Behler}},\ and\ \bibinfo {author}
  {\bibfnamefont {M.}~\bibnamefont {Ceriotti}},\ }\bibfield  {title} {\bibinfo
  {title} {Automatic selection of atomic fingerprints and reference
  configurations for machine-learning potentials},\ }\href
  {https://doi.org/10.1063/1.5024611} {\bibfield  {journal} {\bibinfo
  {journal} {The Journal of Chemical Physics}\ }\textbf {\bibinfo {volume}
  {148}},\ \bibinfo {pages} {241730} (\bibinfo {year} {2018})}\BibitemShut
  {NoStop}%
\bibitem [{\citenamefont {Willatt}\ \emph {et~al.}(2018)\citenamefont
  {Willatt}, \citenamefont {Musil},\ and\ \citenamefont
  {Ceriotti}}]{willatt_feature_2018}%
  \BibitemOpen
  \bibfield  {author} {\bibinfo {author} {\bibfnamefont {M.~J.}\ \bibnamefont
  {Willatt}}, \bibinfo {author} {\bibfnamefont {F.}~\bibnamefont {Musil}},\
  and\ \bibinfo {author} {\bibfnamefont {M.}~\bibnamefont {Ceriotti}},\
  }\bibfield  {title} {\bibinfo {title} {Feature optimization for atomistic
  machine learning yields a data-driven construction of the periodic table of
  the elements},\ }\href {https://doi.org/10.1039/C8CP05921G} {\bibfield
  {journal} {\bibinfo  {journal} {Phys. Chem. Chem. Phys.}\ }\textbf {\bibinfo
  {volume} {20}},\ \bibinfo {pages} {29661} (\bibinfo {year}
  {2018})}\BibitemShut {NoStop}%
\bibitem [{\citenamefont {Darby}\ \emph {et~al.}(2021)\citenamefont {Darby},
  \citenamefont {Kermode},\ and\ \citenamefont
  {Cs{\'a}nyi}}]{darby_compressing_2021}%
  \BibitemOpen
  \bibfield  {author} {\bibinfo {author} {\bibfnamefont {J.~P.}\ \bibnamefont
  {Darby}}, \bibinfo {author} {\bibfnamefont {J.~R.}\ \bibnamefont {Kermode}},\
  and\ \bibinfo {author} {\bibfnamefont {G.}~\bibnamefont {Cs{\'a}nyi}},\
  }\bibfield  {title} {\bibinfo {title} {Compressing local atomic neighbourhood
  descriptors},\ }\href {http://arxiv.org/abs/2112.13055} {\bibfield  {journal}
  {\bibinfo  {journal} {ArXiv211213055 Cond-Mat}\ } (\bibinfo {year} {2021})},\
  \Eprint {https://arxiv.org/abs/2112.13055} {arXiv:2112.13055 [cond-mat]}
  \BibitemShut {NoStop}%
\bibitem [{\citenamefont {Bart{\'o}k}\ \emph {et~al.}(2013)\citenamefont
  {Bart{\'o}k}, \citenamefont {Kondor},\ and\ \citenamefont
  {Cs{\'a}nyi}}]{bartok_representing_2013}%
  \BibitemOpen
  \bibfield  {author} {\bibinfo {author} {\bibfnamefont {A.~P.}\ \bibnamefont
  {Bart{\'o}k}}, \bibinfo {author} {\bibfnamefont {R.}~\bibnamefont {Kondor}},\
  and\ \bibinfo {author} {\bibfnamefont {G.}~\bibnamefont {Cs{\'a}nyi}},\
  }\bibfield  {title} {\bibinfo {title} {On representing chemical
  environments},\ }\bibfield  {journal} {\bibinfo  {journal} {Physical Review
  B}\ }\textbf {\bibinfo {volume} {87}},\ \href
  {https://doi.org/10.1103/PhysRevB.87.184115} {10.1103/PhysRevB.87.184115}
  (\bibinfo {year} {2013})\BibitemShut {NoStop}%
\bibitem [{\citenamefont {Yeh}\ \emph {et~al.}(2004)\citenamefont {Yeh},
  \citenamefont {Chen}, \citenamefont {Lin}, \citenamefont {Gan}, \citenamefont
  {Chin}, \citenamefont {Shun}, \citenamefont {Tsau},\ and\ \citenamefont
  {Chang}}]{yeh_nanostructured_2004}%
  \BibitemOpen
  \bibfield  {author} {\bibinfo {author} {\bibfnamefont {J.-W.}\ \bibnamefont
  {Yeh}}, \bibinfo {author} {\bibfnamefont {S.-K.}\ \bibnamefont {Chen}},
  \bibinfo {author} {\bibfnamefont {S.-J.}\ \bibnamefont {Lin}}, \bibinfo
  {author} {\bibfnamefont {J.-Y.}\ \bibnamefont {Gan}}, \bibinfo {author}
  {\bibfnamefont {T.-S.}\ \bibnamefont {Chin}}, \bibinfo {author}
  {\bibfnamefont {T.-T.}\ \bibnamefont {Shun}}, \bibinfo {author}
  {\bibfnamefont {C.-H.}\ \bibnamefont {Tsau}},\ and\ \bibinfo {author}
  {\bibfnamefont {S.-Y.}\ \bibnamefont {Chang}},\ }\bibfield  {title} {\bibinfo
  {title} {Nanostructured {{High-Entropy Alloys}} with {{Multiple Principal
  Elements}}: {{Novel Alloy Design Concepts}} and {{Outcomes}}},\ }\href
  {https://doi.org/10.1002/adem.200300567} {\bibfield  {journal} {\bibinfo
  {journal} {Adv. Eng. Mater.}\ }\textbf {\bibinfo {volume} {6}},\ \bibinfo
  {pages} {299} (\bibinfo {year} {2004})}\BibitemShut {NoStop}%
\bibitem [{\citenamefont {Cantor}\ \emph {et~al.}(2004)\citenamefont {Cantor},
  \citenamefont {Chang}, \citenamefont {Knight},\ and\ \citenamefont
  {Vincent}}]{cantor_microstructural_2004}%
  \BibitemOpen
  \bibfield  {author} {\bibinfo {author} {\bibfnamefont {B.}~\bibnamefont
  {Cantor}}, \bibinfo {author} {\bibfnamefont {I.~T.~H.}\ \bibnamefont
  {Chang}}, \bibinfo {author} {\bibfnamefont {P.}~\bibnamefont {Knight}},\ and\
  \bibinfo {author} {\bibfnamefont {A.~J.~B.}\ \bibnamefont {Vincent}},\
  }\bibfield  {title} {\bibinfo {title} {Microstructural development in
  equiatomic multicomponent alloys},\ }\href
  {https://doi.org/10.1016/j.msea.2003.10.257} {\bibfield  {journal} {\bibinfo
  {journal} {Materials Science and Engineering: A}\ }\textbf {\bibinfo {volume}
  {375-377}},\ \bibinfo {pages} {213} (\bibinfo {year} {2004})}\BibitemShut
  {NoStop}%
\bibitem [{\citenamefont {Senkov}\ \emph {et~al.}(2010)\citenamefont {Senkov},
  \citenamefont {Wilks}, \citenamefont {Miracle}, \citenamefont {Chuang},\ and\
  \citenamefont {Liaw}}]{senkov_refractory_2010}%
  \BibitemOpen
  \bibfield  {author} {\bibinfo {author} {\bibfnamefont {O.~N.}\ \bibnamefont
  {Senkov}}, \bibinfo {author} {\bibfnamefont {G.~B.}\ \bibnamefont {Wilks}},
  \bibinfo {author} {\bibfnamefont {D.~B.}\ \bibnamefont {Miracle}}, \bibinfo
  {author} {\bibfnamefont {C.~P.}\ \bibnamefont {Chuang}},\ and\ \bibinfo
  {author} {\bibfnamefont {P.~K.}\ \bibnamefont {Liaw}},\ }\bibfield  {title}
  {\bibinfo {title} {Refractory high-entropy alloys},\ }\href
  {https://doi.org/10.1016/j.intermet.2010.05.014} {\bibfield  {journal}
  {\bibinfo  {journal} {Intermetallics}\ }\textbf {\bibinfo {volume} {18}},\
  \bibinfo {pages} {1758} (\bibinfo {year} {2010})}\BibitemShut {NoStop}%
\bibitem [{\citenamefont {Tsai}\ and\ \citenamefont
  {Yeh}(2014)}]{tsai_high-entropy_2014}%
  \BibitemOpen
  \bibfield  {author} {\bibinfo {author} {\bibfnamefont {M.-H.}\ \bibnamefont
  {Tsai}}\ and\ \bibinfo {author} {\bibfnamefont {J.-W.}\ \bibnamefont {Yeh}},\
  }\bibfield  {title} {\bibinfo {title} {High-{{Entropy Alloys}}: {{A Critical
  Review}}},\ }\href {https://doi.org/10.1080/21663831.2014.912690} {\bibfield
  {journal} {\bibinfo  {journal} {Materials Research Letters}\ }\textbf
  {\bibinfo {volume} {2}},\ \bibinfo {pages} {107} (\bibinfo {year}
  {2014})}\BibitemShut {NoStop}%
\bibitem [{\citenamefont {Glielmo}\ \emph {et~al.}(2018)\citenamefont
  {Glielmo}, \citenamefont {Zeni},\ and\ \citenamefont
  {De~Vita}}]{glielmo_efficient_2018}%
  \BibitemOpen
  \bibfield  {author} {\bibinfo {author} {\bibfnamefont {A.}~\bibnamefont
  {Glielmo}}, \bibinfo {author} {\bibfnamefont {C.}~\bibnamefont {Zeni}},\ and\
  \bibinfo {author} {\bibfnamefont {A.}~\bibnamefont {De~Vita}},\ }\bibfield
  {title} {\bibinfo {title} {Efficient nonparametric n -body force fields from
  machine learning},\ }\bibfield  {journal} {\bibinfo  {journal} {Physical
  Review B}\ }\textbf {\bibinfo {volume} {97}},\ \href
  {https://doi.org/10.1103/PhysRevB.97.184307} {10.1103/PhysRevB.97.184307}
  (\bibinfo {year} {2018})\BibitemShut {NoStop}%
\bibitem [{\citenamefont {Vandermause}\ \emph {et~al.}(2020)\citenamefont
  {Vandermause}, \citenamefont {Torrisi}, \citenamefont {Batzner},
  \citenamefont {Xie}, \citenamefont {Sun}, \citenamefont {Kolpak},\ and\
  \citenamefont {Kozinsky}}]{vandermause_--fly_2020}%
  \BibitemOpen
  \bibfield  {author} {\bibinfo {author} {\bibfnamefont {J.}~\bibnamefont
  {Vandermause}}, \bibinfo {author} {\bibfnamefont {S.~B.}\ \bibnamefont
  {Torrisi}}, \bibinfo {author} {\bibfnamefont {S.}~\bibnamefont {Batzner}},
  \bibinfo {author} {\bibfnamefont {Y.}~\bibnamefont {Xie}}, \bibinfo {author}
  {\bibfnamefont {L.}~\bibnamefont {Sun}}, \bibinfo {author} {\bibfnamefont
  {A.~M.}\ \bibnamefont {Kolpak}},\ and\ \bibinfo {author} {\bibfnamefont
  {B.}~\bibnamefont {Kozinsky}},\ }\bibfield  {title} {\bibinfo {title}
  {On-the-fly active learning of interpretable {{Bayesian}} force fields for
  atomistic rare events},\ }\href {https://doi.org/10.1038/s41524-020-0283-z}
  {\bibfield  {journal} {\bibinfo  {journal} {npj Comput Mater}\ }\textbf
  {\bibinfo {volume} {6}},\ \bibinfo {pages} {1} (\bibinfo {year}
  {2020})}\BibitemShut {NoStop}%
\bibitem [{QUI()}]{QUIP}%
  \BibitemOpen
  \href {https://github.com/libAtoms/QUIP} {\bibinfo {title} {{QUIP} -
  {QUantum} mechanics and {Interatomic} {Potentials}}},\ \bibinfo {note}
  {\url{https://github.com/libAtoms/QUIP}}\BibitemShut {NoStop}%
\bibitem [{\citenamefont {Byggm{\"a}star}\ \emph {et~al.}(2019)\citenamefont
  {Byggm{\"a}star}, \citenamefont {Hamedani}, \citenamefont {Nordlund},\ and\
  \citenamefont {Djurabekova}}]{byggmastar_machine-learning_2019}%
  \BibitemOpen
  \bibfield  {author} {\bibinfo {author} {\bibfnamefont {J.}~\bibnamefont
  {Byggm{\"a}star}}, \bibinfo {author} {\bibfnamefont {A.}~\bibnamefont
  {Hamedani}}, \bibinfo {author} {\bibfnamefont {K.}~\bibnamefont {Nordlund}},\
  and\ \bibinfo {author} {\bibfnamefont {F.}~\bibnamefont {Djurabekova}},\
  }\bibfield  {title} {\bibinfo {title} {Machine-learning interatomic potential
  for radiation damage and defects in tungsten},\ }\href
  {https://doi.org/10.1103/PhysRevB.100.144105} {\bibfield  {journal} {\bibinfo
   {journal} {Phys. Rev. B}\ }\textbf {\bibinfo {volume} {100}},\ \bibinfo
  {pages} {144105} (\bibinfo {year} {2019})}\BibitemShut {NoStop}%
\bibitem [{\citenamefont {Bart{\'o}k}\ and\ \citenamefont
  {Cs{\'a}nyi}(2015)}]{bartok_gaussian_2015}%
  \BibitemOpen
  \bibfield  {author} {\bibinfo {author} {\bibfnamefont {A.~P.}\ \bibnamefont
  {Bart{\'o}k}}\ and\ \bibinfo {author} {\bibfnamefont {G.}~\bibnamefont
  {Cs{\'a}nyi}},\ }\bibfield  {title} {\bibinfo {title} {Gaussian approximation
  potentials: {{A}} brief tutorial introduction},\ }\href
  {https://doi.org/10.1002/qua.24927} {\bibfield  {journal} {\bibinfo
  {journal} {International Journal of Quantum Chemistry}\ }\textbf {\bibinfo
  {volume} {115}},\ \bibinfo {pages} {1051} (\bibinfo {year}
  {2015})}\BibitemShut {NoStop}%
\bibitem [{\citenamefont {Bart{\'o}k}\ \emph {et~al.}(2018)\citenamefont
  {Bart{\'o}k}, \citenamefont {Kermode}, \citenamefont {Bernstein},\ and\
  \citenamefont {Cs{\'a}nyi}}]{bartok_machine_2018}%
  \BibitemOpen
  \bibfield  {author} {\bibinfo {author} {\bibfnamefont {A.~P.}\ \bibnamefont
  {Bart{\'o}k}}, \bibinfo {author} {\bibfnamefont {J.}~\bibnamefont {Kermode}},
  \bibinfo {author} {\bibfnamefont {N.}~\bibnamefont {Bernstein}},\ and\
  \bibinfo {author} {\bibfnamefont {G.}~\bibnamefont {Cs{\'a}nyi}},\ }\bibfield
   {title} {\bibinfo {title} {Machine {{Learning}} a {{General}}-{{Purpose
  Interatomic Potential}} for {{Silicon}}},\ }\bibfield  {journal} {\bibinfo
  {journal} {Physical Review X}\ }\textbf {\bibinfo {volume} {8}},\ \href
  {https://doi.org/10.1103/PhysRevX.8.041048} {10.1103/PhysRevX.8.041048}
  (\bibinfo {year} {2018})\BibitemShut {NoStop}%
\bibitem [{\citenamefont {Daw}\ and\ \citenamefont
  {Baskes}(1984)}]{daw_embedded-atom_1984}%
  \BibitemOpen
  \bibfield  {author} {\bibinfo {author} {\bibfnamefont {M.~S.}\ \bibnamefont
  {Daw}}\ and\ \bibinfo {author} {\bibfnamefont {M.~I.}\ \bibnamefont
  {Baskes}},\ }\bibfield  {title} {\bibinfo {title} {Embedded-atom method:
  {{Derivation}} and application to impurities, surfaces, and other defects in
  metals},\ }\href {https://doi.org/10.1103/PhysRevB.29.6443} {\bibfield
  {journal} {\bibinfo  {journal} {Phys. Rev. B}\ }\textbf {\bibinfo {volume}
  {29}},\ \bibinfo {pages} {6443} (\bibinfo {year} {1984})}\BibitemShut
  {NoStop}%
\bibitem [{\citenamefont {Finnis}\ and\ \citenamefont
  {Sinclair}(1984)}]{finnis_simple_1984}%
  \BibitemOpen
  \bibfield  {author} {\bibinfo {author} {\bibfnamefont {M.~W.}\ \bibnamefont
  {Finnis}}\ and\ \bibinfo {author} {\bibfnamefont {J.~E.}\ \bibnamefont
  {Sinclair}},\ }\bibfield  {title} {\bibinfo {title} {A simple empirical
  {{N}}-body potential for transition metals},\ }\href
  {https://doi.org/10.1080/01418618408244210} {\bibfield  {journal} {\bibinfo
  {journal} {Philosophical Magazine A}\ }\textbf {\bibinfo {volume} {50}},\
  \bibinfo {pages} {45} (\bibinfo {year} {1984})}\BibitemShut {NoStop}%
\bibitem [{\citenamefont {Byggm{\"a}star}\ \emph
  {et~al.}(2022{\natexlab{a}})\citenamefont {Byggm{\"a}star}, \citenamefont
  {Nikoulis}, \citenamefont {Fellman}, \citenamefont {Granberg}, \citenamefont
  {Djurabekova},\ and\ \citenamefont {Nordlund}}]{byggmastar_multiscale_2022}%
  \BibitemOpen
  \bibfield  {author} {\bibinfo {author} {\bibfnamefont {J.}~\bibnamefont
  {Byggm{\"a}star}}, \bibinfo {author} {\bibfnamefont {G.}~\bibnamefont
  {Nikoulis}}, \bibinfo {author} {\bibfnamefont {A.}~\bibnamefont {Fellman}},
  \bibinfo {author} {\bibfnamefont {F.}~\bibnamefont {Granberg}}, \bibinfo
  {author} {\bibfnamefont {F.}~\bibnamefont {Djurabekova}},\ and\ \bibinfo
  {author} {\bibfnamefont {K.}~\bibnamefont {Nordlund}},\ }\bibfield  {title}
  {\bibinfo {title} {Multiscale machine-learning interatomic potentials for
  ferromagnetic and liquid iron},\ }\href
  {https://doi.org/10.1088/1361-648X/ac6f39} {\bibfield  {journal} {\bibinfo
  {journal} {J. Phys.: Condens. Matter}\ }\textbf {\bibinfo {volume} {34}},\
  \bibinfo {pages} {305402} (\bibinfo {year} {2022}{\natexlab{a}})}\BibitemShut
  {NoStop}%
\bibitem [{\citenamefont {Zeni}(2020)}]{zeni_gaussian_2020}%
  \BibitemOpen
  \bibfield  {author} {\bibinfo {author} {\bibfnamefont {C.}~\bibnamefont
  {Zeni}},\ }\href@noop {} {\bibinfo {title} {Gaussian process regression for
  nonparametric force fields}} (\bibinfo {year} {2020}),\ \bibinfo {note}
  {{PhD} thesis}\BibitemShut {NoStop}%
\bibitem [{\citenamefont {Mendelev}\ \emph {et~al.}(2003)\citenamefont
  {Mendelev}, \citenamefont {Han}, \citenamefont {Srolovitz}, \citenamefont
  {Ackland}, \citenamefont {Sun},\ and\ \citenamefont
  {Asta}}]{mendelev_development_2003}%
  \BibitemOpen
  \bibfield  {author} {\bibinfo {author} {\bibfnamefont {M.~I.}\ \bibnamefont
  {Mendelev}}, \bibinfo {author} {\bibfnamefont {S.}~\bibnamefont {Han}},
  \bibinfo {author} {\bibfnamefont {D.~J.}\ \bibnamefont {Srolovitz}}, \bibinfo
  {author} {\bibfnamefont {G.~J.}\ \bibnamefont {Ackland}}, \bibinfo {author}
  {\bibfnamefont {D.~Y.}\ \bibnamefont {Sun}},\ and\ \bibinfo {author}
  {\bibfnamefont {M.}~\bibnamefont {Asta}},\ }\bibfield  {title} {\bibinfo
  {title} {Development of new interatomic potentials appropriate for
  crystalline and liquid iron},\ }\href
  {https://doi.org/10.1080/14786430310001613264} {\bibfield  {journal}
  {\bibinfo  {journal} {Philosophical Magazine}\ }\textbf {\bibinfo {volume}
  {83}},\ \bibinfo {pages} {3977} (\bibinfo {year} {2003})}\BibitemShut
  {NoStop}%
\bibitem [{\citenamefont {Ziegler}\ \emph {et~al.}(1985)\citenamefont
  {Ziegler}, \citenamefont {Biersack},\ and\ \citenamefont
  {Littmarck}}]{ziegler_stopping_1985}%
  \BibitemOpen
  \bibfield  {author} {\bibinfo {author} {\bibfnamefont {J.~F.}\ \bibnamefont
  {Ziegler}}, \bibinfo {author} {\bibfnamefont {J.~P.}\ \bibnamefont
  {Biersack}},\ and\ \bibinfo {author} {\bibfnamefont {U.}~\bibnamefont
  {Littmarck}},\ }\bibfield  {title} {\bibinfo {title} {The {{Stopping}} and
  {{Range}} of {{Ions}} in {{Matter}}},\ }in\ \href
  {http://link.springer.com/chapter/10.1007/978-1-4615-8103-1\\_3} {\emph
  {\bibinfo {booktitle} {Treatise on {{Heavy}}-{{Ion Science}}}}}\ (\bibinfo
  {publisher} {{Pergamon}},\ \bibinfo {address} {{New York}},\ \bibinfo {year}
  {1985})\ pp.\ \bibinfo {pages} {93--129}\BibitemShut {NoStop}%
\bibitem [{\citenamefont {Nordlund}\ \emph {et~al.}(1997)\citenamefont
  {Nordlund}, \citenamefont {Runeberg},\ and\ \citenamefont
  {Sundholm}}]{nordlund_repulsive_1997}%
  \BibitemOpen
  \bibfield  {author} {\bibinfo {author} {\bibfnamefont {K.}~\bibnamefont
  {Nordlund}}, \bibinfo {author} {\bibfnamefont {N.}~\bibnamefont {Runeberg}},\
  and\ \bibinfo {author} {\bibfnamefont {D.}~\bibnamefont {Sundholm}},\
  }\bibfield  {title} {\bibinfo {title} {Repulsive interatomic potentials
  calculated using {{Hartree}}-{{Fock}} and density-functional theory
  methods},\ }\href {https://doi.org/10.1016/S0168-583X(97)00447-3} {\bibfield
  {journal} {\bibinfo  {journal} {Nuclear Instruments and Methods in Physics
  Research Section B: Beam Interactions with Materials and Atoms}\ }\textbf
  {\bibinfo {volume} {132}},\ \bibinfo {pages} {45} (\bibinfo {year}
  {1997})}\BibitemShut {NoStop}%
\bibitem [{\citenamefont {Foiles}\ \emph {et~al.}(1986)\citenamefont {Foiles},
  \citenamefont {Baskes},\ and\ \citenamefont
  {Daw}}]{foiles_embedded-atom-method_1986}%
  \BibitemOpen
  \bibfield  {author} {\bibinfo {author} {\bibfnamefont {S.~M.}\ \bibnamefont
  {Foiles}}, \bibinfo {author} {\bibfnamefont {M.~I.}\ \bibnamefont {Baskes}},\
  and\ \bibinfo {author} {\bibfnamefont {M.~S.}\ \bibnamefont {Daw}},\
  }\bibfield  {title} {\bibinfo {title} {Embedded-atom-method functions for the
  fcc metals {{Cu}}, {{Ag}}, {{Au}}, {{Ni}}, {{Pd}}, {{Pt}}, and their
  alloys},\ }\href {https://doi.org/10.1103/PhysRevB.33.7983} {\bibfield
  {journal} {\bibinfo  {journal} {Phys. Rev. B}\ }\textbf {\bibinfo {volume}
  {33}},\ \bibinfo {pages} {7983} (\bibinfo {year} {1986})}\BibitemShut
  {NoStop}%
\bibitem [{\citenamefont {Ackland}\ \emph {et~al.}(2004)\citenamefont
  {Ackland}, \citenamefont {Mendelev}, \citenamefont {Srolovitz}, \citenamefont
  {Han},\ and\ \citenamefont {Barashev}}]{ackland_development_2004}%
  \BibitemOpen
  \bibfield  {author} {\bibinfo {author} {\bibfnamefont {G.~J.}\ \bibnamefont
  {Ackland}}, \bibinfo {author} {\bibfnamefont {M.~I.}\ \bibnamefont
  {Mendelev}}, \bibinfo {author} {\bibfnamefont {D.~J.}\ \bibnamefont
  {Srolovitz}}, \bibinfo {author} {\bibfnamefont {S.}~\bibnamefont {Han}},\
  and\ \bibinfo {author} {\bibfnamefont {A.~V.}\ \bibnamefont {Barashev}},\
  }\bibfield  {title} {\bibinfo {title} {Development of an interatomic
  potential for phosphorus impurities in {$\alpha$}-iron},\ }\href
  {https://doi.org/10.1088/0953-8984/16/27/003} {\bibfield  {journal} {\bibinfo
   {journal} {J. Phys.: Condens. Matter}\ }\textbf {\bibinfo {volume} {16}},\
  \bibinfo {pages} {S2629} (\bibinfo {year} {2004})}\BibitemShut {NoStop}%
\bibitem [{\citenamefont {Ackland}\ and\ \citenamefont
  {Reed}(2003)}]{ackland_two-band_2003}%
  \BibitemOpen
  \bibfield  {author} {\bibinfo {author} {\bibfnamefont {G.~J.}\ \bibnamefont
  {Ackland}}\ and\ \bibinfo {author} {\bibfnamefont {S.~K.}\ \bibnamefont
  {Reed}},\ }\bibfield  {title} {\bibinfo {title} {Two-band second moment model
  and an interatomic potential for caesium},\ }\href
  {https://doi.org/10.1103/PhysRevB.67.174108} {\bibfield  {journal} {\bibinfo
  {journal} {Phys. Rev. B}\ }\textbf {\bibinfo {volume} {67}},\ \bibinfo
  {pages} {174108} (\bibinfo {year} {2003})}\BibitemShut {NoStop}%
\bibitem [{\citenamefont {Thompson}\ \emph {et~al.}(2022)\citenamefont
  {Thompson}, \citenamefont {Aktulga}, \citenamefont {Berger}, \citenamefont
  {Bolintineanu}, \citenamefont {Brown}, \citenamefont {Crozier}, \citenamefont
  {{in 't Veld}}, \citenamefont {Kohlmeyer}, \citenamefont {Moore},
  \citenamefont {Nguyen}, \citenamefont {Shan}, \citenamefont {Stevens},
  \citenamefont {Tranchida}, \citenamefont {Trott},\ and\ \citenamefont
  {Plimpton}}]{thompson_lammps_2022}%
  \BibitemOpen
  \bibfield  {author} {\bibinfo {author} {\bibfnamefont {A.~P.}\ \bibnamefont
  {Thompson}}, \bibinfo {author} {\bibfnamefont {H.~M.}\ \bibnamefont
  {Aktulga}}, \bibinfo {author} {\bibfnamefont {R.}~\bibnamefont {Berger}},
  \bibinfo {author} {\bibfnamefont {D.~S.}\ \bibnamefont {Bolintineanu}},
  \bibinfo {author} {\bibfnamefont {W.~M.}\ \bibnamefont {Brown}}, \bibinfo
  {author} {\bibfnamefont {P.~S.}\ \bibnamefont {Crozier}}, \bibinfo {author}
  {\bibfnamefont {P.~J.}\ \bibnamefont {{in 't Veld}}}, \bibinfo {author}
  {\bibfnamefont {A.}~\bibnamefont {Kohlmeyer}}, \bibinfo {author}
  {\bibfnamefont {S.~G.}\ \bibnamefont {Moore}}, \bibinfo {author}
  {\bibfnamefont {T.~D.}\ \bibnamefont {Nguyen}}, \bibinfo {author}
  {\bibfnamefont {R.}~\bibnamefont {Shan}}, \bibinfo {author} {\bibfnamefont
  {M.~J.}\ \bibnamefont {Stevens}}, \bibinfo {author} {\bibfnamefont
  {J.}~\bibnamefont {Tranchida}}, \bibinfo {author} {\bibfnamefont
  {C.}~\bibnamefont {Trott}},\ and\ \bibinfo {author} {\bibfnamefont {S.~J.}\
  \bibnamefont {Plimpton}},\ }\bibfield  {title} {\bibinfo {title} {{{LAMMPS}}
  - a flexible simulation tool for particle-based materials modeling at the
  atomic, meso, and continuum scales},\ }\href
  {https://doi.org/10.1016/j.cpc.2021.108171} {\bibfield  {journal} {\bibinfo
  {journal} {Computer Physics Communications}\ }\textbf {\bibinfo {volume}
  {271}},\ \bibinfo {pages} {108171} (\bibinfo {year} {2022})}\BibitemShut
  {NoStop}%
\bibitem [{tab()}]{tabgap}%
  \BibitemOpen
  \href {https://gitlab.com/jezper/tabgap} {}\bibinfo {note}
  {\url{https://gitlab.com/jezper/tabgap}}\BibitemShut {NoStop}%
\bibitem [{\citenamefont {Kresse}\ and\ \citenamefont
  {Furthm{\"u}ller}(1996)}]{kresse_efficient_1996}%
  \BibitemOpen
  \bibfield  {author} {\bibinfo {author} {\bibfnamefont {G.}~\bibnamefont
  {Kresse}}\ and\ \bibinfo {author} {\bibfnamefont {J.}~\bibnamefont
  {Furthm{\"u}ller}},\ }\bibfield  {title} {\bibinfo {title} {Efficient
  iterative schemes for ab initio total-energy calculations using a plane-wave
  basis set},\ }\href {https://doi.org/10.1103/PhysRevB.54.11169} {\bibfield
  {journal} {\bibinfo  {journal} {Phys. Rev. B}\ }\textbf {\bibinfo {volume}
  {54}},\ \bibinfo {pages} {11169} (\bibinfo {year} {1996})}\BibitemShut
  {NoStop}%
\bibitem [{\citenamefont {Byggm{\"a}star}\ \emph {et~al.}(2020)\citenamefont
  {Byggm{\"a}star}, \citenamefont {Nordlund},\ and\ \citenamefont
  {Djurabekova}}]{byggmastar_gaussian_2020}%
  \BibitemOpen
  \bibfield  {author} {\bibinfo {author} {\bibfnamefont {J.}~\bibnamefont
  {Byggm{\"a}star}}, \bibinfo {author} {\bibfnamefont {K.}~\bibnamefont
  {Nordlund}},\ and\ \bibinfo {author} {\bibfnamefont {F.}~\bibnamefont
  {Djurabekova}},\ }\bibfield  {title} {\bibinfo {title} {Gaussian
  approximation potentials for body-centered-cubic transition metals},\ }\href
  {https://doi.org/10.1103/PhysRevMaterials.4.093802} {\bibfield  {journal}
  {\bibinfo  {journal} {Phys. Rev. Materials}\ }\textbf {\bibinfo {volume}
  {4}},\ \bibinfo {pages} {093802} (\bibinfo {year} {2020})}\BibitemShut
  {NoStop}%
\bibitem [{\citenamefont {Byggm{\"a}star}\ \emph
  {et~al.}(2022{\natexlab{b}})\citenamefont {Byggm{\"a}star}, \citenamefont
  {Nordlund},\ and\ \citenamefont {Djurabekova}}]{byggmastar_ida_data_2022}%
  \BibitemOpen
  \bibfield  {author} {\bibinfo {author} {\bibfnamefont {J.}~\bibnamefont
  {Byggm{\"a}star}}, \bibinfo {author} {\bibfnamefont {K.}~\bibnamefont
  {Nordlund}},\ and\ \bibinfo {author} {\bibfnamefont {F.}~\bibnamefont
  {Djurabekova}},\ }\href
  {https://doi.org/10.23729/1e6d0215-d26b-4f5f-8f5b-df575efa6594} {\bibinfo
  {title} {Data and {tabGAP} v.2 potential files for {Mo-Nb-Ta-V-W} alloys}}
  (\bibinfo {year} {2022}{\natexlab{b}}),\ \bibinfo {note}
  {https://doi.org/10.23729/1e6d0215-d26b-4f5f-8f5b-df575efa6594}\BibitemShut
  {NoStop}%
\bibitem [{\citenamefont {Lee}\ and\ \citenamefont
  {Baskes}(2000)}]{lee_second_2000}%
  \BibitemOpen
  \bibfield  {author} {\bibinfo {author} {\bibfnamefont {B.-J.}\ \bibnamefont
  {Lee}}\ and\ \bibinfo {author} {\bibfnamefont {M.~I.}\ \bibnamefont
  {Baskes}},\ }\bibfield  {title} {\bibinfo {title} {Second nearest-neighbor
  modified embedded-atom-method potential},\ }\href
  {https://doi.org/10.1103/PhysRevB.62.8564} {\bibfield  {journal} {\bibinfo
  {journal} {Phys. Rev. B}\ }\textbf {\bibinfo {volume} {62}},\ \bibinfo
  {pages} {8564} (\bibinfo {year} {2000})}\BibitemShut {NoStop}%
\bibitem [{\citenamefont {Tersoff}(1988)}]{tersoff_new_1988}%
  \BibitemOpen
  \bibfield  {author} {\bibinfo {author} {\bibfnamefont {J.}~\bibnamefont
  {Tersoff}},\ }\bibfield  {title} {\bibinfo {title} {New empirical approach
  for the structure and energy of covalent systems},\ }\href
  {https://doi.org/10.1103/PhysRevB.37.6991} {\bibfield  {journal} {\bibinfo
  {journal} {Phys. Rev. B}\ }\textbf {\bibinfo {volume} {37}},\ \bibinfo
  {pages} {6991} (\bibinfo {year} {1988})}\BibitemShut {NoStop}%
\bibitem [{\citenamefont {Kauppala}(2021)}]{kauppala_interatomic_2021}%
  \BibitemOpen
  \bibfield  {author} {\bibinfo {author} {\bibfnamefont {J.}~\bibnamefont
  {Kauppala}},\ }\href {http://urn.fi/URN:NBN:fi:hulib-202103021553} {\bibinfo
  {title} {Interatomic potential for molecular dynamics simulations of
  radiation effects in {WMoTa} concentrated alloys}} (\bibinfo {year} {2021}),\
  \bibinfo {note} {{MSc} thesis,
  \url{http://urn.fi/URN:NBN:fi:hulib-202103021553}}\BibitemShut {NoStop}%
\bibitem [{\citenamefont {Morris}\ \emph {et~al.}(1994)\citenamefont {Morris},
  \citenamefont {Wang}, \citenamefont {Ho},\ and\ \citenamefont
  {Chan}}]{Mor94}%
  \BibitemOpen
  \bibfield  {author} {\bibinfo {author} {\bibfnamefont {J.~R.}\ \bibnamefont
  {Morris}}, \bibinfo {author} {\bibfnamefont {C.~Z.}\ \bibnamefont {Wang}},
  \bibinfo {author} {\bibfnamefont {K.~M.}\ \bibnamefont {Ho}},\ and\ \bibinfo
  {author} {\bibfnamefont {C.~T.}\ \bibnamefont {Chan}},\ }\bibfield  {title}
  {\bibinfo {title} {{Melting line of aluminum from simulations of coexisting
  phases}},\ }\href@noop {} {\bibfield  {journal} {\bibinfo  {journal} {Phys.
  Rev. B}\ }\textbf {\bibinfo {volume} {49}},\ \bibinfo {pages} {3109}
  (\bibinfo {year} {1994})}\BibitemShut {NoStop}%
\bibitem [{\citenamefont {Rumble}(2019)}]{rumble_crc_2019}%
  \BibitemOpen
  \bibinfo {editor} {\bibfnamefont {J.}~\bibnamefont {Rumble}},\ ed.,\
  \href@noop {} {\emph {\bibinfo {title} {{{CRC Handbook}} of {{Chemistry}} and
  {{Physics}}, 100th {{Edition}}}}},\ \bibinfo {edition} {100th}\ ed.\
  (\bibinfo  {publisher} {{CRC Press}},\ \bibinfo {address} {{Boca Raton,
  Fla.}},\ \bibinfo {year} {2019})\BibitemShut {NoStop}%
\bibitem [{\citenamefont {Kresse}\ and\ \citenamefont
  {Hafner}(1993)}]{kresse_ab_1993}%
  \BibitemOpen
  \bibfield  {author} {\bibinfo {author} {\bibfnamefont {G.}~\bibnamefont
  {Kresse}}\ and\ \bibinfo {author} {\bibfnamefont {J.}~\bibnamefont
  {Hafner}},\ }\bibfield  {title} {\bibinfo {title} {Ab initio molecular
  dynamics for liquid metals},\ }\href
  {https://doi.org/10.1103/PhysRevB.47.558} {\bibfield  {journal} {\bibinfo
  {journal} {Phys. Rev. B}\ }\textbf {\bibinfo {volume} {47}},\ \bibinfo
  {pages} {558} (\bibinfo {year} {1993})}\BibitemShut {NoStop}%
\bibitem [{\citenamefont {Perdew}\ \emph {et~al.}(1996)\citenamefont {Perdew},
  \citenamefont {Burke},\ and\ \citenamefont
  {Ernzerhof}}]{perdew_generalized_1996}%
  \BibitemOpen
  \bibfield  {author} {\bibinfo {author} {\bibfnamefont {J.~P.}\ \bibnamefont
  {Perdew}}, \bibinfo {author} {\bibfnamefont {K.}~\bibnamefont {Burke}},\ and\
  \bibinfo {author} {\bibfnamefont {M.}~\bibnamefont {Ernzerhof}},\ }\bibfield
  {title} {\bibinfo {title} {Generalized {{Gradient Approximation Made
  Simple}}},\ }\href {https://doi.org/10.1103/PhysRevLett.77.3865} {\bibfield
  {journal} {\bibinfo  {journal} {Phys. Rev. Lett.}\ }\textbf {\bibinfo
  {volume} {77}},\ \bibinfo {pages} {3865} (\bibinfo {year}
  {1996})}\BibitemShut {NoStop}%
\end{thebibliography}%

\end{document}